\documentclass[12pt,a4paper]{article}
\usepackage{amsmath,amsthm,amsfonts,amssymb,bbm}
\usepackage{graphicx,psfrag}
\usepackage{cite}

\numberwithin{equation}{section}

\renewcommand{\Re}{\mathrm{Re}}

\newcommand{\tpng}{{\mathbf{t}}}
\newcommand{\xpng}{{\mathbf{x}}}
\newcommand{\Or}{\mathcal{O}}
\newcommand{\Ai}{\mathrm{Ai}}
\newcommand{\Pb}{\mathbbm{P}}
\newcommand{\Id}{\mathbbm{1}}
\newcommand{\e}{\varepsilon}
\newcommand{\I}{{\rm i}}
\newcommand{\R}{\mathbb{R}}
\newcommand{\N}{\mathbb{N}}
\newcommand{\Z}{\mathbb{Z}}
\newcommand{\dx}{\mathrm{d}}
\newcommand{\Af}{{\cal A}_{\rm 1}}

\newcommand{\Nc}{{\cal N}}

\newtheorem{prop}{Proposition}
\newtheorem{thm}[prop]{Theorem}
\newtheorem{lem}[prop]{Lemma}
\newtheorem{defin}[prop]{Definition}

\newenvironment{proofOF}[2]{\removelastskip\vspace{6pt}\noindent {\it Proof of #1.}~\rm#2}{\qed \par\vspace{6pt}}

\title{Large time asymptotics of growth models on space-like paths II: PNG and parallel TASEP}
\author{Alexei Borodin\thanks{California Institute of Technology, e-mail: borodin@caltech.edu},
Patrik L. Ferrari\thanks{Weierstrass Institute (WIAS), Berlin, e-mail: ferrari@wias-berlin.de},\\
Tomohiro Sasamoto\thanks{Chiba University, e-mail: sasamoto@math.s.chiba-u.ac.jp}}

\date{July 27, 2007}

\begin{document}
\maketitle \sloppy

\begin{abstract}
We consider the polynuclear growth (PNG) model in 1+1 dimension with
flat initial condition and no extra constraints. The joint
distributions of surface height at finitely many points at a fixed
time moment are given as marginals of a signed determinantal point
process. The long time scaling limit of the surface height is shown
to coincide with the Airy$_1$ process. This result holds more
generally for the observation points located along any space-like
path in the space-time plane. We also obtain the corresponding
results for the discrete time TASEP (totally asymmetric simple
exclusion process) with parallel update.
\end{abstract}

\section{Introduction} \label{SectIntro}
\begin{figure}[t!]
\begin{center}
\psfrag{x}{$x$}
\psfrag{h}{$h(x,t)$}
\includegraphics[height=2.5cm]{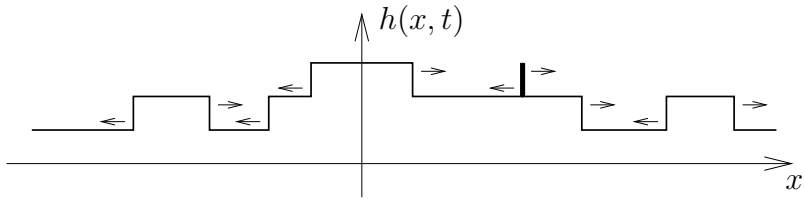}
\caption{Illustration of the PNG height and its dynamics. The bold vertical piece is a nucleation. The arrows indicate the movements of the steps. A Java animation of the PNG dynamics is available at~\cite{FerPNG}.}
\label{FigPNG}
\end{center}
\end{figure}

The main focus of this work is a stochastic growth model in $1+1$ dimensions, called the polynuclear growth (PNG) model. It belongs to the KPZ (Kardar-Parisi-Zhang~\cite{KPZ86}) universality class and it can be described as follows (see Figure~\ref{FigPNG}). At time $t$, the surface is described by an integer-valued height function $x\mapsto h(x,t)\in \Z$, $x\in \R,t\in \R_+$. It thus consists of up-steps ($\lrcorner\hspace{-0.15em}\ulcorner$) and down-steps ($\urcorner\hspace{-0.15em}\llcorner$). The dynamics has a deterministic and a stochastic part:\\[0.5em]
(a) up- (down-) steps move to the left (right) with unit speed and disappear upon colliding,\\[0.5em]
(b) pairs of up- and down- steps (nucleations) are randomly added on the surface with some given intensity.\\[0.5em]
The up- and down-steps of the nucleations then spread out with unit
speed according to (a). The PNG model can be interpreted in several
different ways, see~\cite{FP05} for a review.

On a macroscopic scale the surface of the PNG model grows
deterministically, i.e., $\lim_{t\to\infty}t^{-1} h(\xi t,t)=H(\xi)$
is a non-random function. However, on a mesoscopic scale
fluctuations grow in time. This is called roughening in statistical
physics and extensive numerical studies have been made~\cite{BS95}.
Since the PNG model is in the KPZ universality class, the
fluctuation of the surface height is expected to live on a $t^{1/3}$
scale and non-trivial correlations are to be seen on a $t^{2/3}$
scale. Therefore, to have an interesting large time limit, we have
to rescale the surface height as
\begin{equation}\label{eqFirst}
\frac{h(u t^{2/3},t)-t H(u t^{-1/3})}{t^{1/3}}.
\end{equation}

One of the most natural initial conditions for PNG is the flat
initial condition, i.e., $h(x,0) = 0$ for all $x\in\R$. We consider
nucleations occurring with translation-invariant
intensity.\footnote{In other words, the nucleation events form a
Poisson process with constant intensity in the space-time upper
half-plane.} We refer to the PNG model with such initial condition
as \emph{flat PNG}. In this case, by mapping the flat PNG to a
point-to-line directed percolation model it was
proven~\cite{BR99,PS00,PS00b} that the one-point distribution is, in
the $t\to\infty$ limit, the GOE Tracy-Widom distribution $F_1$,
first discovered in random matrix theory~\cite{TW96}. However, no
information on joint height distributions at more than one point has
been previously known.

\newpage

\vspace{0.5em} \noindent \emph{New Results.} The main results of
this paper are precisely the computation and asymptotic analysis of
these joint distributions. In particular, we prove the convergence
of the height rescaled as in (\ref{eqFirst}) to the Airy$_1$ process
in the $t\to\infty$ limit (see Section~\ref{subSectScalingLimit} for
a definition of the process). The Airy$_1$ process has been
discovered in the context of the asymmetric exclusion
process~\cite{Sas05,BFPS06,BFP06,BFS07}. Our result, stated in
Theorem~\ref{ThmXiTAi}, is obtained by first determining an
expression for the joint distributions for finite time $t$
(Proposition~\ref{PropPNG}) and then taking the appropriate scaling
limit.

Proposition~\ref{PropPNG} is actually just a particular case of
Theorem~\ref{ThmcPNG}, where we determine joint distributions along
any space-like paths (as in Minkowski diagram), for which fixed time
is a special case. The scaling limit is analyzed at this level of
generality, thus Theorem~\ref{ThmXiTAi} holds for any space-like
paths. In contrast to previous works on the subject, our approach
does not rely on the so-called RSK correspondence (RSK for
Robinson-Schensted-Knuth), which was successfully applied for corner
growth models, but does not seem to be well suited for the flat
growth.

On the way of getting the results for the flat PNG, we consider a
discrete time version of it, the Gates-Westcott
dynamics~\cite{GW95, PS02}.
This model is closely related to the totally
asymmetric simple exclusion process (TASEP) in discrete time with
parallel update and alternating initial conditions. The
corresponding results for this model are Theorem~\ref{ThmKalt} for
the joint distributions along space-like paths, and
Theorem~\ref{ThmXitoAi} for the convergence to the Airy$_1$ process
in the scaling limit. For the TASEP, the extreme situations of
space-like paths are positions of different particles at a fixed
time and positions of a fixed particle at different time moments
(tagged particle). The space-like extension for TASEP is based on
the previous paper~\cite{BF07}.

\vspace{0.5em} \noindent \emph{Previous works on PNG.} Another type
of initial conditions for the PNG model has been analyzed before. It
is the corner growth geometry, where nucleations occur only inside
the cone $\{|x|\leq t\}$. The limit shape $H$ is a semi-circle, and
the model is called \emph{PNG droplet}. In this geometry, the limit
process has been obtained in~\cite{PS02}; it is known as Airy$_2$
process (previously called simply Airy process). The approach uses
an extension to a multilayer model (inherited from the RSK
construction), see ~\cite{PS02,Jo02b}. The multilayer method was
also used in other related models~\cite{Jo03b,Jo03,BO04,SI03,IS04a,SI04b}.
Also, for the flat PNG it was used to connect the associated point
process at a single position and the point process of GOE
eigenvalues~\cite{Fer04}. Results on the behavior for the PNG droplet
along space-like paths can be found in \cite{BO04}. For a very brief
description of the previously known results on TASEP fluctuations
see the introductions of \cite{BF07,Sas07}.

\subsubsection*{Outline}
In section \ref{SectModel}, we introduce our models and state the
results. In section \ref{SectJointDistr}, we give an expression of
the transition probability of the discrete TASEP as a marginal of a
determinantal signed point process. In section
\ref{SectJointSpaceLike} the Fredholm determinant expression for the
joint distributions is obtained. The argument substantially relies
on the algebraic techniques of \cite{BF07}. In section
\ref{SectPOTXitoAi}, we consider the scaling limit of the parallel
TASEP. In section \ref{SectPOTcPNG}, the continuous time PNG model
is considered. In section \ref{SectPOTXiTAi}, we consider the
scaling limit for the continous PNG model.

\subsubsection*{Acknowledgments}
A. Borodin was partially supported by the NSF grants DMS-0402047 and
DMS-0707163. P.L. Ferrari is grateful to H. Spohn for useful
discussions. The work of T. Sasamoto is supported by the
Grant-in-Aid for Young Scientists (B), the Ministry of Education,
Culture, Sports, Science and Technology, Japan.

\section{Models and results} \label{SectModel}
We start from the discrete time TASEP with parallel update. Then we will make the connection with a discrete version of the PNG, from which the continuous time PNG is obtained.

\subsection{Discrete time TASEP with parallel update}\label{subSectFiniteSyst}
We consider discrete time TASEP with parallel update and alternating initial conditions, i.e., particle $i$ has initial position $x_i(0)=-2i$, $i\in\Z$. At each time step, each particle hops to its right neighbor site with probability $p=1-q$ provided that
the site is empty. The particle positions at time $t$ is denoted by
$x_i(t), i\in\Z$.

The dynamics of a particle depends only on particles on its right. This fact allows us to determine the joint distributions of particle positions also for different times, but restricted to "space-like paths". To define what we mean with "space-like paths", we consider a sequence of couples $(n_i,t_i)$, where $n_i$ is the number of the particle and $t_i$ is the time when this particle is observed. On such couples we define a partial order $\prec$, given by
\begin{equation}\label{DefPrec}
(n_i,t_i)\prec (n_j,t_j)\textrm{ if }n_j\geq n_i, t_j\leq t_i, \textrm{ and the two couples are not identical.}
\end{equation}
A space-like path is a sequence of ordered couples, namely,
\begin{equation}
{\cal S}=\{(n_k,t_k),k=1,2,\ldots| (n_k,t_k)\prec (n_{k+1},t_{k+1})\}.
\end{equation}
The reason of the name "space-like" will be clear in the large
time limit, where everything becomes continuous. Then space-like
is the same concept as in the Minkowski diagram. The border cases
for space-like paths are fixed time ($t_i\equiv t, \forall i$)
and fixed particle number ($n_i \equiv n, \forall i$).
The next theorem concerns the joint distributions of particle positions.

\begin{thm}\label{ThmKalt}
Let particle with label $i$ start at $x_i(0)=-2i$, $i\in\Z$. Consider a space-like path $\cal S$. For any given $m$, the joint distribution of the positions of the first $m$ points in $\cal S$ is given by
\begin{equation}
\Pb\Big(\bigcap_{k=1}^m \big\{x_{n_k}(t_k) \geq a_k\big\}\Big)
=\det(\Id-\chi_a^{(-)}K\chi_a^{(-)})_{\ell^2(\{(n_1,t_1),\ldots,(n_m,t_m)\}\times\Z)}
\end{equation}
where $\chi_a^{(-)}((n_k,t_k),x)=\Id(x<a_k)$. The kernel $K_t$ is given by
\begin{eqnarray}\label{K}
K((n_1,t_1),x_1;(n_2,t_2),x_2)& =& -\phi^{((n_1,t_1),(n_2,t_2))}(x_1,x_2) \Id_{[(n_1,t_1)\prec (n_2,t_2)]}\nonumber \\ &+&\widetilde{K}((n_1,t_1),x_1;(n_2,t_2),x_2),
\end{eqnarray}
where
\begin{eqnarray}\label{Kt}
& & \widetilde{K}((n_1,t_1),x_1;(n_2,t_2),x_2)\nonumber \\
&=&\frac{-1}{2\pi\I}\oint_{\Gamma_0} \dx z
\frac{(1+z)^{x_2+n_1+n_2}}{(-z)^{x_1+n_1+n_2+1}}
\frac{(1-p)^{t_1-2n_1-x_1}}{(1+pz)^{t_1+t_2+1-(x_1+n_1+n_2)}},
\end{eqnarray}
and
\begin{eqnarray}\label{phi}
& & \phi^{((n_1,t_1),(n_2,t_2))}(x_1,x_2)\nonumber \\
&=&\frac{1}{2\pi\I}\oint_{\Gamma_{-1}}\dx w \frac{(1+pw)^{t_1-t_2}}{(1+w)^{x_1-x_2+1}} \left(\frac{-w}{(1+w)(1+pw)}\right)^{n_1-n_2}
\end{eqnarray}
where $\Gamma_0$ (resp.\ $\Gamma_{-1}$) is any simple loop, anticlockwise oriented, with $0$ (resp.\ $-1$) being the unique pole of the integrand inside the contour.
\end{thm}
\textbf{Remark.} In the limit $p\to 0$ under the time scaling by $p^{-1}$ the discrete time TASEP converges to the continuous time TASEP, and Theorem~\ref{ThmKalt} turns into a special case of Proposition 3.6 of~\cite{BF07}, where a more general continuous time model called PushASEP was considered.

\subsection{Airy$_1$ process and scaling limit}\label{subSectScalingLimit}
Starting from Theorem~\ref{ThmKalt} we can analyze large time limits. The limit process is the so-called Airy$_1$ process introduced in \cite{Sas05,BFPS06}, which we recall here.
\begin{defin}[The Airy$_1$ process]\label{Defin}
Define the extended kernel,
\begin{eqnarray}\label{eqKernelExpanded}
& &\hspace{-2em}
K_{\Af}(\tau_1,\xi_1;\tau_2,\xi_2)
=-\frac{1}{\sqrt{4\pi (\tau_2-\tau_1)}}
 \exp\left(-\frac{(\xi_2-\xi_1)^2}{4 (\tau_2-\tau_1)}\right)
 \Id(\tau_2>\tau_1) \nonumber \\
& &\hspace{-2em}
+ \Ai(\xi_1+\xi_2+(\tau_2-\tau_1)^2)
  \exp \left((\tau_2-\tau_1)(\xi_1+\xi_2)+\frac23(\tau_2-\tau_1)^3\right).
\end{eqnarray}
The \emph{Airy$_1$ process}, $\Af$, is the process with $m$-point joint distributions at $\tau_1<\tau_2<\ldots<\tau_m$ given by the Fredholm determinant,
\begin{equation}\label{eqTransAiryProcess}
\Pb\bigg(\bigcap_{k=1}^m\{\Af(\tau_k)\leq s_k\}\bigg)
=\det(\Id-\chi_s K_{\Af} \chi_s)_{L^2(\{\tau_1,\ldots,\tau_m\}\times\R)},
\end{equation}
where $\chi_s(\tau_k,x)=\Id(x>s_k)$.
\end{defin}

\begin{figure}[t!]
\begin{center}
\psfrag{n}{$n$}
\psfrag{t}{$t$}
\psfrag{w1}{$\theta$}
\psfrag{f}[c]{$\pi(\theta)$}
\includegraphics[height=4cm]{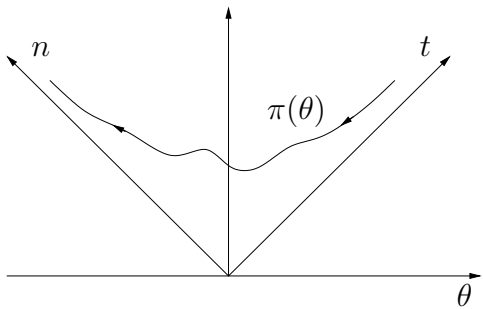}
\caption{An example of a space-like path $\pi(\theta)$. Its slope is, in absolute value, at most $1$.}
\label{FigSpaceLikePath}
\end{center}
\end{figure}

Theorem~\ref{ThmKalt} allows us to analyze joint distributions of particle positions for situations spanning between fixed time and fixed particle number (the tagged particle problem). One way to parametrize such situations is via a space-like path. We thus consider an arbitrary smooth function $\pi$ satisfying
\begin{equation}
|\pi'(\theta)| \leq 1\textrm{ and } \pi(\theta)+\theta>0,
\end{equation}
see Figure~\ref{FigSpaceLikePath}. The requirement $\pi(\theta)+\theta>0$ reflects $t>0$.
Then, we choose couples of $(t,n)$ on
$\{((\pi(\theta)+\theta)T,(\pi(\theta)-\theta)T),\theta\in\R\}$, where
$T$ is a large parameter. The case of fixed time, say $t=T$, is obtained
by setting $\pi(\theta)=1-\theta$, while fixed particle number, say
$n=\alpha T$, by $\pi(\theta)=\alpha+\theta$ with some constant $\alpha$.

From KPZ scaling exponents~\cite{KPZ86},
we expect to see a nontrivial limit if
we consider positions at distance of order $T^{2/3}$. Thus, the
focus on the region around $\theta T$ is given by $\theta T-u T^{2/3}$, i.e.,
setting $\theta-u T^{-1/3}$ instead of $\theta$ and, by series
expansions, we scale time and particle number as
\begin{eqnarray}\label{eqScaling1}
t(u)&=&\lfloor(\pi(\theta)+\theta) T -(\pi'(\theta)+1) u T^{2/3} + \tfrac12 \pi''(\theta) u^2 T^{1/3}\rfloor,\nonumber \\
n(u)&=&\lfloor(\pi(\theta)-\theta) T +(1-\pi'(\theta)) u T^{2/3} + \tfrac12 \pi''(\theta) u^2 T^{1/3}\rfloor.
\end{eqnarray}
The KPZ fluctuation exponent is $1/3$, thus we expect to see fluctuations of particle positions on a scale of order $T^{1/3}$. Therefore, we define the rescaled process $\Xi_T$ by
\begin{equation}\label{Xit}
u \mapsto \Xi_T(u) = \frac{x_{n(u)}(t(u))-(-2 n(u)+\mathbf{v} t(u))}{-T^{1/3}}.
\end{equation}
Here the mean speed of particles, $\mathbf{v}$, is determined to
be $\mathbf{v}=1-\sqrt{q}$ from the subsequent asymptotic analysis
but can be known beforehand from the stationary measure for density
$1/2$~\cite{JPS95}.
This process has a limit as $T\to\infty$ given in terms of the Airy$_1$ process.

\begin{thm}\label{ThmXitoAi}
Let $\Xi_T$ be the rescaled process as in (\ref{Xit}). Then
\begin{equation}
 \lim_{T\to\infty} \Xi_T(u) = \kappa_{\rm v}\Af(\kappa_{\rm h} u),
\end{equation}
in the sense of finite dimensional distributions. The vertical (fluctuations) and horizontal (correlations) scaling coefficients are given by
\begin{align}\label{eqkappa0}
 \kappa_{\rm v} &= (\pi(\theta)+\theta)^{1/3}(1-q)^{1/3} q^{1/6}, \\
 \kappa_{\rm h} &= \frac{(\pi(\theta)+\theta)^{2/3}(1-q)^{2/3} q^{-1/6}}{(\pi'(\theta)+1)(1-\sqrt{q})/2+1-\pi'(\theta)}.
\end{align}
\end{thm}
\textbf{Remark.} A similar result for the PushASEP with alternating initial condition has been proved in Theorem 2.2 of~\cite{BF07}.

\subsection{TASEP and growth models}\label{subsectTASEPgrowth}
As mentioned in the introduction, the discrete TASEP with parallel
update is related
to a surface growth model from which the polynuclear growth
model in continuous time can be obtained as a limit.
Let $t\geq 0$ and $x \in\mathbb{R}$ denote the time
and the one-dimensional space coordinate respectively, and let
$h_t(x)$ be the height of the surface at time $t$ and at position $x$.
Let us introduce a dynamics of $h_t(x)$ as follows.
Initially, at time $t=0$, the surface is flat;
$h_0(x)=0$, for all $x\in \mathbb{R}$.
Right after each integer time $(t=0+,1+,2+,\ldots)$,
there could occur a nucleation with width 0 and height 1
with probability $q$ ($0< q <1$) independently at each integer
position $x$ such that $t+x+h_t(x)$ is even.
Each nucleation is regarded as consisting of an upstep and a
downstep and each upstep (resp. downstep) moves to the left
(resp. right) with unit speed. This is a deterministic
part of the evolution.
\begin{figure}[t!]
\begin{center}
\psfrag{t=0}{\hspace{-8mm}$t=0$}
\psfrag{t=0+}{\hspace{-8mm}$t=0+$}
\psfrag{t=1/2}{\hspace{-8mm}$t=\frac12$}
\psfrag{t=1}{\hspace{-8mm}$t=1$}
\psfrag{t=1+}{\hspace{-8mm}$t=1+$}
\psfrag{t=3/2}{\hspace{-8mm}$t=\frac32$}
\psfrag{t=2}{\hspace{-8mm}$t=2$}
\psfrag{t=2+}{\hspace{-8mm}$t=2+$}
\psfrag{t=3}{\hspace{-8mm}$t=3$}
\includegraphics[height=10cm]{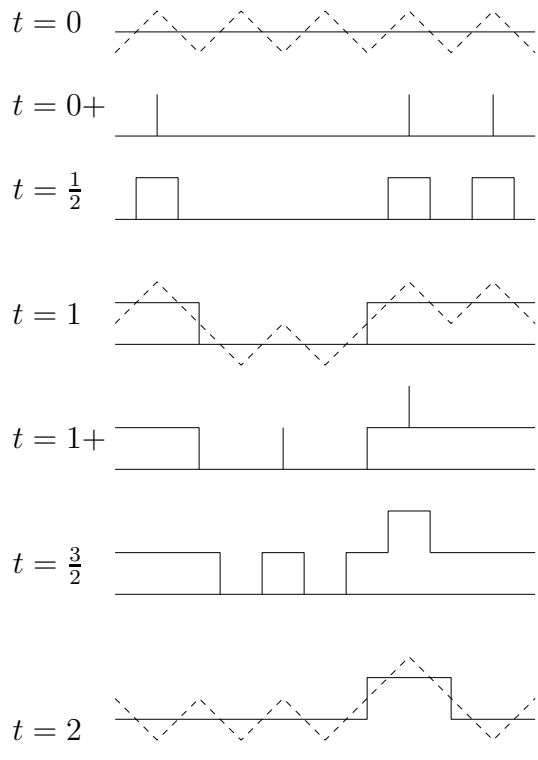}
\caption{A surface growth model. For half-odd integer times
this is equivalent to the discretized Gates-Westcott dynamics
and for integer times to the discrete TASEP.}
\label{FigGW}
\end{center}
\end{figure}
When an upstep and a downstep collide, they merge together.
See the solid line in Figure~\ref{FigGW} for an example until $t=2$.
The dynamics of the growth model, if we focus only on half-odd
times ($t=\frac12, \frac32, \ldots$),
is the same as one considered in \cite{PS02}, i.e., a
discretized version of the Gates-Westcott dynamics~\cite{GW95}.
It is known that in an appropriate $q\to 0$ limit this growth
model reduces to the standard continuous time PNG model
\cite{PS02}.

To see the connection to the discrete TASEP, let us focus on integer
times $(t=0,1,2,\ldots)$ and positions ($x\in\mathbb{Z}$) from now on
and represent the surface as consisting of
elementary upward slopes $\diagup$ and downward slopes $\diagdown$
as indicated by dashed lines in Figure~\ref{FigGW}.
At $t=0$, even (resp. odd) $x$'s are taken to be the center of
the upward (resp. downward) slopes.
Then the dynamics of the surface is described as follows: At each
time step the surface grows upward by unit height deterministically
and then each local maximum ($\diagup\diagdown$) of slope turns
into a local minimum ($\diagdown\diagup$) independently
with probability $p\equiv 1-q$. If we interpret an upward (resp. a downward)
slope as a site occupied by a particle (resp. an empty site), this is
equivalent to the discrete time TASEP with parallel update
under the alternating initial condition.
\begin{figure}[t!]
\begin{center}
\psfrag{x}{$x$}
\psfrag{h}{$h$}
\psfrag{n}{$n$}
\includegraphics[height=5cm]{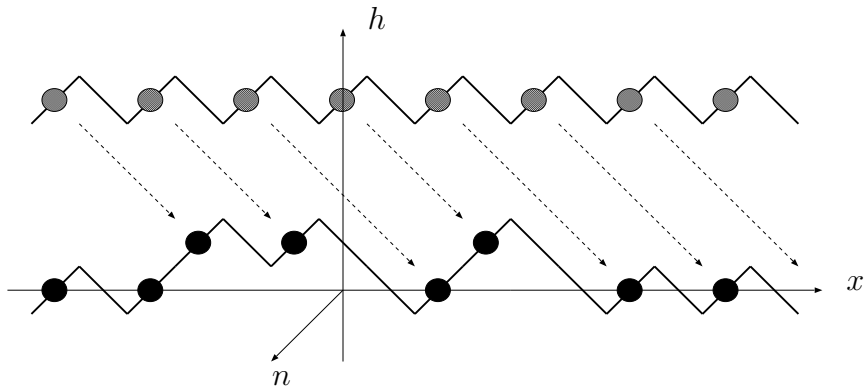}
\caption{Surface height and TASEP particle positions.
An expample for $t=4$.}
\label{FigGWTASEP}
\end{center}
\end{figure}

The relation between the surface height $h_t(x)$ and the
position of the TASEP particle is given by
\begin{equation}\label{hx}
 h_t(x) \leq H \Leftrightarrow x_{\lfloor\frac{t-x-H}{2}\rfloor}(t) \geq x
\end{equation}
and is understood as follows. On the plot of the surface at some
fixed time $t$, draw also the initial surface at $h=t$. See Figure~\ref{FigGWTASEP} for an
expample. Then, from the correspondence between the growth model
and the TASEP, the surface at time $t$ can be regarded as the
particle positions. In this plot particles move along the down-right
direction as indicated. The left hand side of (\ref{hx}) is
equivalent to the condition that the TASEP particle corresponding
to $(x,h=t-H)$ has already reached $x$. Since the axis of the particle
number $n$ is in the down-left direction, the value of $n$
corresponding to $(x,h=t-H)$ is $\lfloor (t-H-x)/2 \rfloor$.
This consideration results in the relation (\ref{hx}).
From the relation (\ref{hx}) the joint distributions of the
height of the growth model is readily obtained through
\begin{equation}
\Pb\Big(\bigcap_{i=1}^m\{h_{t_i}(x_i)\leq H_i\}\Big)
=
\Pb\Big(\bigcap_{i=1}^m\{x_{n_i}(t_i)\geq x_i\}\Big),
\label{hxn}
\end{equation}
combined with Theorem \ref{ThmKalt}.

When $q\to 0$, the TASEP particles move almost deterministically and
the surface $h_t(x)$ grows slowly, when a
particle decides not to jump (with probability $q$).
The continuous time PNG model is obtained by taking $q\to 0$ while
setting space and time units to $\sqrt{q}/2$ (the $2$ is chosen to
have nucleations with intensity $2$ like in \cite{PS02}).
Denote by $\xpng$ and $\tpng$ the position and time variables
in the continuous time
PNG model. The PNG height function $h^{\rm PNG}(\xpng,\tpng)$ is then
obtained by the limit
\begin{equation}\label{eqHeightRescaling}
h^{\rm PNG}(\xpng,\tpng)=\lim_{q\to 0} h_{2\tpng/\sqrt{q}}(-2\xpng/\sqrt{q}).
\end{equation}
Here the minus sign on the right hand side is put for a convenience.
The results below do not depend on this sign because of the
symmetry of the model in consideration. The joint distribution of the surface height
at time $\tpng$ is given as follows.

\begin{prop}\label{PropPNG}
Consider $m$ space positions $\xpng_1<\xpng_2<\ldots<\xpng_m$. Then, the joint distribution at time $\tpng$ of the heights $h^{\rm PNG}(\xpng_k,\tpng)$, $k=1,\ldots,m$, is given by
\begin{equation}
\Pb\Big(\bigcap_{k=1}^m \big\{h^{\rm PNG}(\xpng_k,\tpng) \leq H_k\big\}\Big)
=\det(\Id-\chi_H K_\tpng^{\rm PNG}\chi_H)_{\ell^2(\{\xpng_1,\ldots,\xpng_m\}\times\Z)}
\end{equation}
where the kernel is given by
\begin{align}
& K_\tpng^{\rm PNG}(\xpng_1,h_1;\xpng_2,h_2)
 = -I_{|h_1-h_2|}\left(2(\xpng_2-\xpng_1)\right)\Id(\xpng_2>\xpng_1)  \notag\\
&+ \left(\frac{2\tpng+\xpng_2-\xpng_1}{2\tpng-\xpng_2+\xpng_1}\right)^{(h_1+h_2)/2}
J_{h_1+h_2}\left(2\sqrt{4\tpng^2-(\xpng_2-\xpng_1)^2}\right) \Id(2\tpng \geq |\xpng_2-\xpng_1|)
\end{align}
where $I_n(x)$ and $J_n(x)$ are the modified Bessel functions and the Bessel functions, see e.g.~\cite{AS84}.
\end{prop}
The last indicator function is obvious if one thinks about the PNG
model. In fact, the height at position $\xpng$ at time $\tpng$
depends on events lying in the backward light cone of $(\xpng,\tpng)$
on $\R\times\R_+$. Thus, when $|\xpng_2-\xpng_1|>2\tpng$, the backwards
light cones of $(\xpng_1,\tpng)$ and $(\xpng_2,\tpng)$ do not intersect
in $\R\times\R_+$, which implies that the two height functions are
independent. The Fredholm determinant then splits into blocks.

The result of Proposition~\ref{PropPNG} is actually a specialization
of a more general situation which follows from the TASEP
correspondence. In the TASEP, the space-like paths $\pi$ we had for
particle numbers and times become the paths
\begin{equation}
(\xpng,\tpng)=\big(\pi(\theta)-3\theta,\theta+\pi(\theta)\big).
\end{equation}
The condition $|\pi'(\theta)|\leq 1$ implies that $\partial \tpng / \partial \xpng \in [-1,0]$, i.e., these are space-like paths as in special relativity oriented into the past. By the symmetry of the problem, one can consider also space-like paths locally oriented into the future, just looking at the process in the other direction.

Denote by $\gamma$ such a path on $\R\times\R_+^*$, i.e., $(\xpng,\tpng=\gamma(\xpng))$, then $\theta$ and $\pi(\theta)$ are given by the relations
\begin{equation}
\theta=(\gamma(\xpng)-\xpng)/4,\quad \pi(\theta)=(3\gamma(\xpng)+\xpng)/4,
\end{equation}
and the joint distributions of the surface height along the path $\gamma$ are expressed as in Theorem~\ref{ThmcPNG}.

\begin{thm}\label{ThmcPNG}
Let us denote by $\tpng_k=\gamma(\xpng_k)$. Then, the joint distributions of $h^{\rm PNG}(\xpng_k,\tpng_k)$, $k=1,\ldots,m$, is given by
\begin{equation}
\Pb\Big(\bigcap_{k=1}^m \big\{h^{\rm PNG}(\xpng_k,\tpng_k) \leq H_k\big\}\Big)
=\det(\Id-\chi_H K^{\rm PNG}\chi_H)_{\ell^2(\{(\xpng_1,\tpng_1),\ldots,(\xpng_k,\tpng_k)\}\times\Z)}
\end{equation}
where the kernel is given by
\begin{align}
& K^{\rm PNG}((\xpng_1,\tpng_1),h_1;(\xpng_2,\tpng_2),h_2)
= -\left(\frac{\xpng_2-\xpng_1+\tpng_1-\tpng_2}{\xpng_2-\xpng_1-\tpng_1+\tpng_2}\right)^{(h_1-h_2)/2} \notag\\
&\times  I_{|h_1-h_2|}\left(2\sqrt{(\xpng_2-\xpng_1)^2-(\tpng_2-\tpng_1)^2}\right)
\Id_{\{(\tpng_1+\xpng_1,\tpng_1)\prec (\tpng_2+\xpng_2,\tpng_2)\}}  \notag\\
&+ \left(\frac{(\tpng_1+\tpng_2)+(\xpng_2-\xpng_1)}
              {(\tpng_1+\tpng_2)-(\xpng_2-\xpng_1)}\right)^{(h_2+h_1)/2}
 J_{h_1+h_2}\left(2\sqrt{(\tpng_1+\tpng_2)^2-(\xpng_2-\xpng_1)^2}\right)
 \notag\\
&\times \Id(\tpng_1+\tpng_2 \geq |\xpng_1-\xpng_2|)
\label{PNGkernel}
\end{align}
where $I_n(x)$ and $J_n(x)$ are the modified Bessel functions and the Bessel functions.
The condition $\Id_{\{(\tpng_1+\xpng_1,\tpng_1)\prec (\tpng_2+\xpng_2,\tpng_2)\}}$ means that $\xpng_2-\xpng_1\geq \tpng_1-\tpng_2>0$ or $\xpng_2-\xpng_1>\tpng_1-\tpng_2\geq 0$.
\end{thm}

In the first term, for $\xpng_2>\xpng_1$, the condition
$\xpng_2-\xpng_1\geq \tpng_1-\tpng_2$ is satisfied for
$\tpng_k=\gamma(\xpng_k)$. Also, notice that when
$\xpng_2-\xpng_1\to \tpng_1-\tpng_2$, the first term of the
kernel goes to $(2(\xpng_2-\xpng_1))^{|h_1-h_2|}/(|h_1-h_2|)!$ .

\subsection{Scaling limit for the continuous PNG model}\label{subsectScalingPNG}
\begin{figure}[t!]
\begin{center}
\psfrag{x}{$\xpng$}
\psfrag{t}[l]{$\tpng$}
\psfrag{g}[l]{$\tpng=T\gamma(\xpng/T)$}
\includegraphics[height=3.5cm]{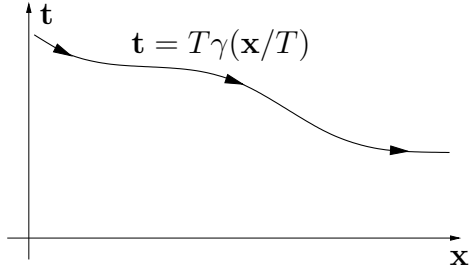}
\caption{A space-time path $\gamma$ for continuous time PNG. $T$ is proportional to the PNG time $\tpng$.}
\label{FigPNGpath}
\end{center}
\end{figure}

The last result of this paper is the large time behavior of the flat PNG. The large parameter denoted by $T$ is proportional to time $\tpng$. Using the function $\gamma$, we consider $\tpng=T \gamma(\xpng/T)$, see Figure~\ref{FigPNGpath}.

Since the system is translation invariant, we focus around the origin, i.e., we look at the PNG height at
\begin{equation}\label{eq1.24}
\begin{cases}
\xpng(u)=u T^{2/3},\\
\tpng(u)=\gamma(0)T+\gamma'(0)u T^{2/3}+\tfrac12 \gamma''(0) u^2 T^{1/3}.
\end{cases}
\end{equation}
The surface height grows with the speed equal to $2$. Thus, for large time $\tpng$, the macroscopic height will be close to $2\tpng$. Fluctuations live on a $T^{1/3}$ scale. Consequently, we define the rescaled height process $h_T^{\rm PNG}$ by
\begin{equation}
u\mapsto h_T^{\rm PNG}(u) = \frac{h^{\rm PNG}(\xpng(u),\tpng(u))-2\tpng(u)}{T^{1/3}}.
\label{XPt}
\end{equation}
The large $T$ (thus large time too) behavior of $h_T^{\rm PNG}$ is given in terms of the Airy$_1$ process as stated below.
\begin{thm}\label{ThmXiTAi}
Let $h_T^{\rm PNG}$ be the rescaled process as in (\ref{XPt}). Then, in the limit of large $T$, we have
\begin{equation}
 \lim_{T\to\infty} h_T^{\rm PNG}(u) = S_v\Af(S_h u),
\end{equation}
in the sense of finite dimensional distributions. The scaling coefficients $S_v$ and $S_h$ are given by
\begin{equation}
S_v= (2\gamma(0))^{1/3},\quad S_h= (2\gamma(0))^{2/3}=S_v^2.
\end{equation}
\end{thm}
\noindent For $\gamma(x)=1$, i.e., fixed time, this was conjectured to hold in~\cite{BFPS06}.

\section{Transition probability for the finite system}\label{SectJointDistr}
Let $G(x_1,\ldots,x_N;t)$ denote the transition probability of the parallel TASEP with $N$ particles starting at $t=0$ at positions $y_N<\ldots<y_1$. This is the probability that the $N$ particles starting from positions $y_N < \ldots < y_1$ at $t=0$ are at positions $x_N < \ldots < x_1$ at $t$.

Consider a determinantal signed point process on the set \mbox{$\underline{x}=\{x_i^n,1\leq i \leq n \leq N\}$} by setting the measure
\begin{equation}\label{WN}
 W_N(\underline{x})
 = \bigg(\prod_{n=1}^{N-1} \det(\phi^{\sharp}(x_i^n,x_{j+1}^{n+1}))_{0\leq i,j\leq n}\bigg) \det(F_{-i+1}(x^N_j-y_{N+1-i},t+1-i))_{1\leq i,j\leq N}
\end{equation}
where
\begin{equation}\label{eq2.3}
 \phi^{\sharp}(x,y) =
 \begin{cases}
  1, & y \geq x, \\
  p, & y= x-1, \\
  0,  & y\leq x-2,
 \end{cases}
\end{equation}
the function $F_n(x,t)$ defined by
\begin{equation}
 F_{-n}(x,t)
 = \frac{1}{2\pi\I} \oint_{\Gamma_{0,-1}} \dx w
 \frac{w^n}{(1+w)^{n+x+1}} (1+(1-q)w)^t,
\label{F}
\end{equation}
and where we used the convention, $x_0^n = -\infty$.

The following proposition states that the one time transition probability of the TASEP is given as a marginal of the signed measure (\ref{WN}).
\begin{prop}\label{GP}
Let us set $x_1^n=x_n, n=1,\ldots,N$. Then
\begin{equation}
 G(x_1,\ldots,x_N;t) = \sum_{\mathcal{D}} W_N(\underline{x})
\end{equation}
where summation is over the variables in the set,
\begin{equation}\label{eq3.5}
 \mathcal{D} = \{ x_i^n, 2\leq i \leq n \leq N | x_i^n > x_{i-1}^n \}
\end{equation}
varying over $\Z$.
\end{prop}
Note that $W_N(\underline{x})$ is actually symmetric with respect to permutations of variables with same upper index, so the ordering in (\ref{eq3.5}) is used for singling out the minimal $x_1^n=\min\{x_i^n,i=1,\ldots,n\}$.

\textbf{Remark.} Similar representations for the transition probability of continuous time TASEP, discrete time TASEP with sequential update and PushASEP have been obtained in~\cite{BFPS06,BFP06,BF07}.

In the different parts of the proof of Proposition~\ref{GP}, we will use several properties of the function $F_n$, which are listed below.
\begin{lem} \label{propF}
\begin{align}
 F_{n+1}(x,t) &= \sum_{y=x}^{\infty} F_n(y,t), \\
 F_n(x,t+1) &= q F_n(x,t) + (1-q) F_n(x-1,t) \label{pF1}\\
            &= F_n(x,t) + (1-q) F_{n-1}(x-1,t), \label{pF2}\\
(\phi^{\sharp} * F_n)(x,t) &= F_{n+1}(x,t+1), \label{pF3}\\
 F_{-n}(x,-n) &= 0 \quad \text{{\rm for} } x<-n, n>0, \label{pF4}\\
 F_n(x,n) &= 0 \quad \text{{\rm for} } x>n, n>0, \label{pF5}\\
 F_n(n,n)&=(1-q)^n, n\geq 0 , \label{pF6}\\
 F_{-n}(-n,-n) &=1/(-q)^n, n\geq 0. \label{pF7}
\end{align}
Here ``$*$'' represents the convolution: $(\phi^{\sharp} * f)(x) = \sum_y \phi^{\sharp}(x,y) f(y)$.
\end{lem}

\begin{proofOF}{Lemma \ref{propF}}
These are proven by using the definition (\ref{eq2.3}) and (\ref{F}).
\end{proofOF}

The first step in the proof of Proposition~\ref{GP} is the following Lemma.
\begin{lem} \label{det_f}
Let us set
\begin{equation}
 \phi_{\nu}^{\sharp}(x,y) =
 \begin{cases}
  \nu^{y-x}, & y \geq x, \\
  1-q,       & y= x-1, \\
  0,         & y\leq x-2
 \end{cases}
\end{equation}
and $\phi_{\nu}^{\sharp}(-\infty,y)=\nu^y$.
Then, for any antisymmetric function $f(b_1,\ldots,b_n)$,
\begin{align}
 &\quad
 \sum_{\substack{b_n>\ldots >b_1>b_0\\ b_0:\rm{fixed}}}
 \det(\phi_{\nu}^{\sharp}(a_i,b_j))_{0\leq i,j\leq n}
 \cdot f(b_1,\ldots,b_n) \notag\\
 &=
 g_{\nu}(a_1,b_0)
 \sum_{\substack{b_n>\ldots >b_1>b_0\\ b_0:\rm{ fixed}}}
 \det(\phi_{\nu}^{\sharp}(a_i,b_j))_{1\leq i,j\leq n} \cdot f(b_1,\ldots,b_n)
\label{df_gdf}
\end{align}
where $a_n > \ldots > a_1,a_0=-\infty$ and
\begin{equation}
 g_{\nu}(a,b) = \begin{cases}
       0,                 & b \geq a, \\
           \nu^b(1-(1-q)\nu), & b=a-1, \\
           \nu^b,             & b \leq a-2.
      \end{cases}
\end{equation}
\end{lem}

\begin{proofOF}{Lemma \ref{det_f}}
From the antisymmetry of $f$ and of the determinant, (\ref{df_gdf}) is equivalent to
\begin{align}
 &\quad
 \sum_{\substack{b_1,\ldots,b_n>b_0\\ b_0:\rm{fixed}}}
 \det(\phi_{\nu}^{\sharp}(a_i,b_j))_{0 \leq i,j \leq n}
 \cdot f(b_1,\ldots,b_n) \notag\\
 &=
 g_{\nu}(a_1,b_0)
 \sum_{\substack{b_1,\ldots,b_n>b_0\\ b_0:\rm{fixed}}}
 \det(\phi_{\nu}^{\sharp}(a_i,b_j))_{1\leq i,j\leq n} \cdot f(b_1,\ldots,b_n).
\label{df_gdf2}
\end{align}
Since a basis of the antisymmetric functions is made of the
antisymmetric delta functions and the relation to prove is
linear in $f$, it is enough to consider
\begin{equation}
 f(b_1,\ldots,b_n)
 =
 \begin{cases}
  (-1)^{\sigma}, & \text{if}~ (b_1,\ldots,b_n)
                            =(b_{\sigma_1},\ldots,b_{\sigma_n})~
                   \text{for some}~\sigma \in S_n, \\
  0, & \text{otherwise}
 \end{cases}
\end{equation}
for fixed $b_1,\ldots,b_n>b_0$. Here $S_n$ is the group of all
permutations of $\{1,\ldots,n\}$.
For this special choice of $f$, the left hand side
of (\ref{df_gdf2}) is $n!$ times the single determinant,
\begin{equation}
 \det
 \begin{bmatrix}
  \nu^{b_0} & \nu^{b_1} & \ldots & \nu^{b_n} \\
  \phi_{\nu}^{\sharp}(a_1,b_0) & \phi_{\nu}^{\sharp}(a_1,b_1) & \ldots & \phi_{\nu}^{\sharp}(a_1,b_n) \\
  \vdots & \vdots & & \vdots \\
  \phi_{\nu}^{\sharp}(a_n,b_0) & \phi_{\nu}^{\sharp}(a_n,b_1) & \ldots & \phi_{\nu}^{\sharp}(a_n,b_n)
 \end{bmatrix}.
\end{equation}

We have the following three cases.\\
\vspace{0.5em}
\noindent (a) $a_1\leq b_0$: the second row gives $(\nu^{b_0-a_1},\ldots,\nu^{b_n-a_1})$
which is proportional to the first row. Therefore in this case the LHS is zero.

\vspace{0.5em}
\noindent (b) $a_1=b_0+1$: The second row is $(1-q,\nu^{b_1-a_1},\ldots,\nu^{b_n-a_1})$. Subtracting
$\nu^{a_1}$ times the second row from the first row one obtains
\begin{equation}
 \nu^{b_0}(1-(1-q)\nu)
 \cdot \det(\phi_{\nu}^{\sharp}(a_i,b_j))_{1\leq i,j\leq n}.
\end{equation}

\vspace{0.5em}
\noindent (c) $a_1>b_0+1$: The first column is $(\nu^{b_0},0,\ldots,0)^t$. Thus the determinant is
$\nu^{b_0}\cdot\det(\phi_{\nu}^{\sharp}(a_i,b_j))_{1\leq i,j\leq n}$.

\vspace{0.5em}
\noindent The result in each case agrees with $n!$ times the RHS of (\ref{df_gdf2}) and hence the lemma is proved.
\end{proofOF}

Let $\Nc(x_1,\ldots,x_N)$ denote the number of $j$'s s.t.\ $x_j-x_{j+1}=1, j=1,\ldots,N-1$.
Using the above lemma with $\nu=1$ in which case $\phi_{\nu}^{\sharp}$
reduces to $\phi^{\sharp}$, we have the following result.
\begin{lem} \label{Gtilde}
With $x_1^n=x_n, n=1,\ldots,N$, one has
\begin{eqnarray}\label{Gdet}
 \sum_{\mathcal{D}} W_N(\underline{x})
 &=&  q^{\Nc(x_1,\ldots,x_N)} \det[F_{j-i}(x_{N-j+1}-y_{N-i+1},t+j-i)]_{1\leq i,j\leq N}\nonumber \\
 &=:& \widetilde{G}(x_1,\ldots,x_N;t).
\end{eqnarray}
\end{lem}

\begin{proofOF}{Lemma \ref{Gtilde}}
For simplicity, we denote
\begin{equation}
 f_i(x) = F_{-i+1}(x-y_{N-i+1},t-i+1),
\label{fi}
\end{equation}
for $i=1,\ldots,N$. From the definitions (\ref{WN}), the LHS of (\ref{Gdet}) writes
\begin{equation}
 \sum_{\substack{x_n^n>x_{n-1}^n > \ldots >x_1^n\\ x_1^n:{\rm fixed}, 1\leq n\leq N}}
 \bigg(\prod_{n=1}^{N-1} \det(\phi^{\sharp}(x_i^n,x_{j+1}^{n+1}))_{0\leq i,j\leq n}\bigg) \det(f_i(x_j^N))_{1\leq i,j\leq N}.
\label{phif}
\end{equation}
Applying Lemma~\ref{det_f} with $\nu=1, n=N-1$, $a_i=x_i^{N-1}$, $i=1,\ldots,N-1$,
$b_i=x_{i+1}^N$, $i=0,\ldots,N-1$ and
\begin{align}
 f(b_1,\ldots,b_n) &= \det(f_i(x_j^N))_{1\leq i,j\leq N},
\end{align}
we obtain
\begin{align}
 (\ref{phif})&=
 g_1(x_1^{N-1},x_1^N) \cdot
 \sum_{\substack{x_n^n>x_{n-1}^n > \ldots > x_1^n \\ x_1^n:{\rm fixed}, 1\leq n\leq N-1}}
 \bigg(\prod_{n=1}^{N-2} \det(\phi^{\sharp}(x_i^n,x_{j+1}^{n+1}))_{0\leq i,j\leq n}\bigg)
 \notag\\
 &\quad \times
 \sum_{\substack{x_N^N > x_{N-1}^N > \ldots > x_1^N\\x_1^N:{\rm fixed}}}
 \det(\phi^{\sharp}(x_i^{N-1},x_{j+1}^N))_{1\leq i,j\leq N-1} \cdot
 \det(f_i(x_j^N))_{1\leq i,j \leq N} .
\label{phif1}
\end{align}
Heine's identity,
\begin{equation}
 \frac{1}{n!} \sum_{x_1,\ldots,x_n} \det(\varphi_i(x_j))_{1\leq i,j\leq n}
 \det(\psi_i(x_j))_{1\leq i,j\leq n}
 =
 \det\big[ \phi_i *\psi_j\big]_{1\leq i,j\leq n},
\end{equation}
allows us to rewrite the last summation in (\ref{phif1}) as
\begin{equation}
 \det
 \begin{bmatrix}
  f_1(x_1^N) & (\phi^{\sharp} * f_1)(x_1^{N-1}) & \ldots & (\phi^{\sharp} * f_1)(x_{N-1}^{N-1})\\
  \vdots     & \vdots  & & \vdots \\
  f_N(x_1^N) & (\phi^{\sharp} * f_N)(x_1^{N-1}) & \ldots & (\phi^{\sharp} * f_N)(x_{N-1}^{N-1})
 \end{bmatrix}.
\end{equation}
We repeat the procedure up to a total of $j-1$ times in column $j$ and we get
\begin{equation}
 (\ref{phif1})=\bigg(\prod_{n=1}^{N-1} g_1(x_1^n,x_1^{n+1})\bigg) \det[\overbrace{\phi^{\sharp} * \ldots * \phi^{\sharp}}^{j-1}
     * f_i(x_1^{N-j+1})]_{1\leq i,j\leq N}.
\label{phif2}
\end{equation}
The proof of the lemma is finished using (\ref{fi}), (\ref{pF3})
and $\prod_{n=1}^{N-1} g_1(x_1^n,x_1^{n+1}) = q^{\Nc(x_1,\ldots,x_N)}$.
\end{proofOF}

\begin{proofOF}{Proposition \ref{GP}}
We need to prove
\begin{equation}
G(x_1,\ldots,x_N;t)=\widetilde{G}(x_1,\ldots,x_N;t).
\label{GG}
\end{equation}
This statement was also proved in~\cite{PP06} by the Bethe ansatz techniques.
Our proof is by induction in $t$. We start by showing that the initial conditions agree, i.e., $\widetilde{G}(x_1,\ldots,x_N;0)=G(x_1,\ldots,x_N;0)$, that is,
\begin{equation}
 q^{\Nc(x_1^1,\ldots,x_1^N)}\cdot
 \det[F_{j-i}(x_{N-j+1}-y_{N-i+1},j-i)]_{1\leq i,j\leq N}
 =
 \prod_{n=1}^N \delta_{x_n,y_n}.
\label{IC}
\end{equation}
We first show that LHS of (\ref{IC}) is zero if $x_N\neq y_N$. If $x_N\leq y_N-1$, since $y_{N-i+1}\geq y_N+i-1$, one has $x_N-y_{N-i+1}<-i+1, i=1,\ldots,N$. Then, from (\ref{pF4}) we have $F_{1-i}(x_N-y_{N-i+1},1-i)=0$, i.e., the first column of LHS of (\ref{IC}) is zero.
Similarly, if $x_N\geq y_N+1$, since $x_{N-j+1} \geq x_N+j-1$, one has $x_{N-j+1}-y_N>j-1, j=1,\ldots,N$. Then, from
(\ref{pF5}) we have $F_{j-1}(x_{N-j+1}-y_N,j-1)=0$, i.e., the first row of LHS of (\ref{IC}) is zero.
This agrees with RHS of (\ref{IC}) also being zero if $x_N\neq y_N$.

Now let us assume $x_N=y_N$. There are two cases.

\noindent (a) $y_{N-1}>y_N+1$. In this case, since $x_N-y_{N-i+1}=y_N-y_{N-i+1}<-i+1, i=2,\ldots,N$, one has $F_{1-i}(x_N-y_{N-i+1},1-i)=0, i=2,\ldots,N$. Then the first column of LHS of (\ref{IC}) is $(1,0,\ldots,0)^t$ and
hence the determinant is equal to $\det[F_{j-i}(x_{N-j+1}-y_{N-i+1},j-i)]_{2\leq i,j\leq N}$.

\vspace{0.5em}
\noindent (b) $y_{N-1}=y_N+1$. First let us see that LHS of (\ref{IC}) is zero when $x_{N-1}\neq y_{N-1}$. We have $x_{N-1} \geq x_N+1 = y_N+1=y_{N-1}$. If $x_{N-1}\geq y_{N-1}+1$, we have $x_{N-j+1}-y_N \geq x_{N-1}+j-2-(y_{N-1}-1)\geq j$, for $j=2,\ldots,N$, and $x_{N-j+1}-y_{N-1}\geq j-1$, for $j=2,\ldots,N$. Then the first and the second row of LHS of (\ref{IC}) are both of the form, $(*,0,\ldots,0)$ where $*$ represents
an arbitrary number and hence the determinant is zero. Hence LHS of (\ref{IC}) is zero if $x_{N-1}\neq y_{N-1}$. On the other hand, when $x_{N-1}=y_{N-1}$, the upper-left $2\times 2$ submatrix of the determinant is
\begin{equation}
 \begin{bmatrix}
  F_0(0,0)      & F_1(1,1) \\
  F_{-1}(-1,-1) & F_0(0,0)
 \end{bmatrix}
 =
 \begin{bmatrix}
  1 & 1-q \\
  -1/q & 1
 \end{bmatrix},
\end{equation}
whose determinant is $1/q$.

Repeating the same procedure, at each step one has either case (a) or (b). The final result is that $y_k=x_k$, for $k=1,\ldots,N$, otherwise the determinant in LHS of (\ref{IC}) is zero. Moreover, when $y_k=x_k$, $k=1,\ldots,N$,
denote by $n_1,n_1+n_2,\ldots,n_1+\ldots+n_\ell$ the values of $j$ such that $x_{j-1}-x_j>1$. Then LHS of (\ref{IC}) is equal to $\prod_{m=1}^\ell D_{n_m}$ with
\begin{equation}
 D_n = \det\left[F_{j-i}(j-i,j-i)\right]_{1\leq i,j\leq n}
\end{equation}
Finally using (\ref{pF6}), (\ref{pF7}), we obtain an explicit form of the matrix. To compute its determinant it is enough to develop along the first row. The determinant of the $(1,1)$ minor is $D_{n-1}$, while the one of the $(1,2)$ minor is $(-1/q) D_{n-1}$ because the minor is the same as the $(1,1)$ minor except the first column is multiplied by $-1/q$. All the other minors have determinant zero, because the first two column are linearly dependent. Thus, $D_n=1\cdot D_{n-1}-(1-q)/(-q) D_{n-1}$, and since $D_1=1$, it follows that
\begin{equation}
 D_n = \frac{1}{q^{n-1}}.
\end{equation}
This ends the part of the proof concerning initial conditions.

Next we prove that (\ref{GG}) holds for $t+1$ if it does for $t$. Since this is true for $t=0$, by induction it will be true for all $t\in\N$. $G$ satisfies the TASEP dynamics, thus
\begin{eqnarray}\label{Gw}
& & G(x_1,\ldots,x_N;t+1)  \nonumber \\
&=& \sum_z G(z_1,\ldots,z_N,t) w(z,x) =\sum_z \widetilde{G}(z_1,\ldots,z_N,t) w(z,x) \\
&=& \sum_z w(z,x)q^{\Nc(z_1,\ldots,z_N)} \det[F_{j-i}(z_{N-j+1}-y_{N-i+1},t+j-i)]_{1\leq i,j\leq N} .\nonumber
\end{eqnarray}
Here
\begin{align}
 w(z,x) = \prod_{n=1}^N v_n, \quad
 v_n    = \begin{cases}
        1,  & z_n = z_{n-1}-1, x_n = z_n, \\
            q,  & z_n < z_{n-1}-1, x_n = z_n, \\
           1-q, & z_n < z_{n-1}-1, x_n = z_n+1,
       \end{cases}
\end{align}
and in the second equality we have used the assumption of
the induction. We rewrite $\widetilde{G}(x_1,\ldots,x_N;t+1)$ using (\ref{pF1}) and (\ref{pF2}) as follows. For $k$ from $1$ to $N$:\\[0.5em]
(a) if $x_k=x_{k+1}+1$, then we use (\ref{pF2}) to the $N+1-k$th column. Then, the new term with the $(1-q)$ factor in front cancels out because it is proportional to its left column of the determinant.\\[0.5em]
(b) if $x_k>x_{k+1}+1$, then we just use (\ref{pF1}).\\[0.5em]
With these replacements we get
\begin{eqnarray}\label{Gtw}
& & \widetilde{G}(x_1,\ldots,x_N;t+1)  \\
&=& \sum_z  \tilde{w}(z,x) q^{\Nc(x_1,\ldots,x_N)} \det[F_{j-i}(z_{N-j+1}-y_{N-i+1},t+j-i)]_{1\leq i,j\leq N},\nonumber
\end{eqnarray}
where
\begin{align}
 \tilde{w}(z,x) = \prod_{n=1}^N \tilde{v}_n, \quad
 \tilde{v}_n    =
           \begin{cases}
        1,  & x_n = x_{n+1}+1, z_n = x_n, \\
            q,  & x_n > x_{n+1}+1, z_n = x_n, \\
           1-q, & x_n > x_{n+1}+1, z_n = x_n-1.
       \end{cases}
\end{align}
\begin{figure}[t!]
\begin{center}
\psfrag{(a)}[c]{$(a)$}
\psfrag{(b)}[c]{$(b)$}
\psfrag{(c)}[c]{$(c)$}
\psfrag{(d)}[c]{$(d)$}
\psfrag{t}[c]{$t$}
\psfrag{t+1}[c]{$t+1$}
\psfrag{q}[b][B][0.8]{$q$}
\psfrag{1}[b][B][0.8]{$1$}
\psfrag{q1}[b][B][0.8]{$q-1$}
\psfrag{qm1}[c][][0.8]{$q^{m-1}$}
\psfrag{qm2}[c][][0.8]{$q^{m-2}$}
\includegraphics[width=14cm]{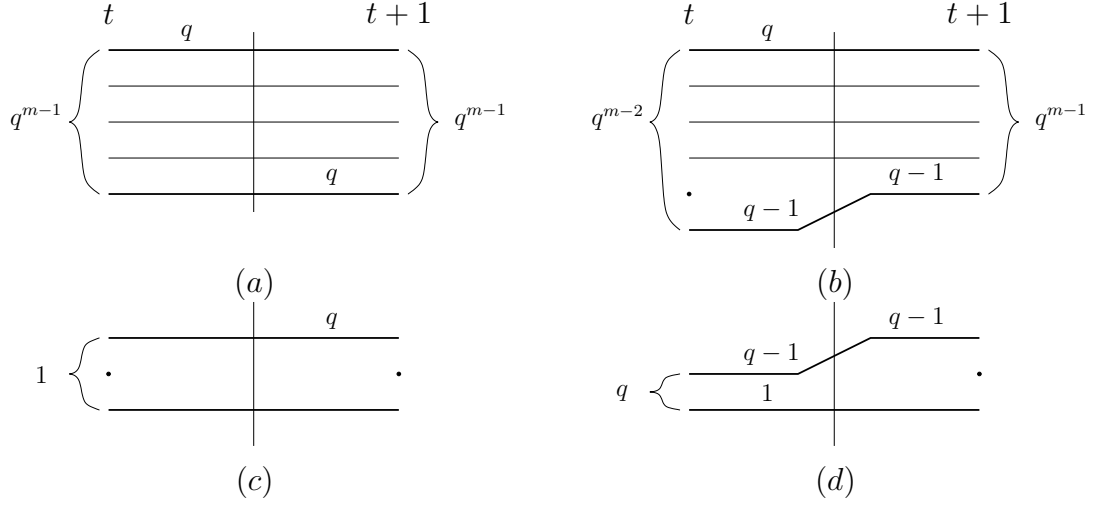}
\caption{Graphical representation of (\ref{eq2.38}). The dots represents empty places, while a line leaving/arriving to a point is an occupied position. In (a) and (b), on the left (resp.\ right) we indicate the weights different from $1$ of LHS (resp. RHS) of (\ref{eq2.38}). In (c) and (d) the bottom and top lines of two blocks at distance $2$ at time $t+1$ are represented, for the cases corresponding to (a) and (b) for the top block.}
\label{FigIC}
\end{center}
\end{figure}
Comparing (\ref{Gw}) and (\ref{Gtw}), it is enough to show
\begin{equation}\label{eq2.38}
q^{\Nc(z_1,\ldots,z_N)} w(z,x) = q^{\Nc(x_1,\ldots,x_N)} \tilde{w}(z,x).
\end{equation}
This indeed holds and can be seen by checking case by case. We illustrate it using Figure~\ref{FigIC}.
First consider a block of particles, say $m$ of them at time $t+1$. There are two possibility of reaching this situations in one time step, as indicated in Figure~\ref{FigIC} (a) and (b). The products of all the weights on the right and on the left are the same, i.e., (\ref{eq2.38}) holds for a single block of particles. If two blocks of particles at time $t+1$ are at distance at least $2$, they are independent during one time step. We just have to check that (\ref{eq2.38}) holds for two blocks at distance $2$ at time $t+1$. Case (a) is illustrated in (c) and the weights are unchanged for both blocks. Case (b) is illustrated in (d). This time, the $q$ on the top line of the second block becomes a $1$, but this is compensated by an extra factor $q$ on the left.
\end{proofOF}

\section{Joint distributions along space-like paths}\label{SectJointSpaceLike}
\begin{thm}\label{ThmJointCorr}
Let us consider particles starting from $y_1>y_2>\ldots$ and denote $x_j(t)$ the position of $j$th particle at time $t$. Consider a sequences of particles and times which are space-like, i.e., a sequence of $m$ such couples ${\cal S}=\{(n_k,t_k),k=1,\ldots,m| (n_k,t_k)\prec (n_{k+1},t_{k+1})\}$. The joint distribution of their positions $x_{n_k}(t_k)$ is given by
\begin{equation}
\Pb\Big(\bigcap_{k=1}^m \big\{x_{n_k}(t_k) \geq a_k\big\}\Big)=
\det(\Id-\chi_a K\chi_a)_{\ell^2(\{(n_1,t_1),\ldots,(n_m,t_m)\}\times\Z)}
\end{equation}
where $\chi_a((n_k,t_k),x)=\Id(x<a_k)$. Here $K$ is the extended kernel with
entries
\begin{equation}\label{eqKernelFinal}
K((n_1,t_1),x_1;(n_2,t_2),x_2) = -\phi^{*((n_1,t_1),(n_2,t_2))}(x_1,x_2)+\sum_{k=1}^{n_2} \Psi^{n_1,t_1}_{n_1-k}(x_1) \Phi^{n_2,t_2}_{n_2-k}(x_2)
\end{equation}
where
\begin{eqnarray}\label{eq2.5}
& &\phi^{*((n_1,t_1),(n_2,t_2))}(x_1,x_2)\\
 &=& \frac{1}{2\pi\I}\oint_{\Gamma_{0,-1}}\dx w \frac{(1+pw)^{t_1-t_2}}{(1+w)^{x_1-x_2+1}} \left(\frac{w}{(1+w)(1+pw)}\right)^{n_1-n_2}\Id_{[(n_1,t_1)\prec (n_2,t_2)]},\nonumber
\end{eqnarray}
and the functions $\Psi^{n,t}_{n-l}$ are given by
\begin{equation}\label{eq2.4}
\Psi^{n,t}_{n-k}(x)=\frac{1}{2\pi\I}\oint_{\Gamma_{0,-1}}\dx w \frac{(1+pw)^t}{(1+w)^{x-y_k+1}}\left(\frac{w}{(w+1)(1+pw)}\right)^{n-k}.
\end{equation}
The functions $\Phi^{n,t}_{n-k}$ are determined by the two conditions
\begin{equation}\label{ortho}
\sum_{x\in\Z}\Psi^{n,t}_{n-l}(x) \Phi^{n,t}_{n-k}(x) = \delta_{k,l},\quad 1\leq k,l\leq n,
\end{equation}
and
\begin{equation}
{\rm span}\{\Phi^{n,t}_{n-k}(x),k=1,\ldots,n\}={\rm span}\{1,x,\ldots,x^{n-1}\}.
\end{equation}
The paths $\Gamma_{0,-1}$ are any simple loops anticlockwise oriented including $0$, $-1$ and no other poles.
\end{thm}

\begin{proofOF}{Theorem~\ref{ThmJointCorr}}
This is the analogue of Proposition 3.1 of~\cite{BF07}. The first step to obtain this proposition, was Lemma 4.4 of~\cite{BF07}. In the simple case where in the model studied in~\cite{BF07} all particles have the same jump rates, i.e., $v_i=1$ for all $i$, then Lemma 4.4 of~\cite{BF07} and our Proposition~\ref{GP} have exactly the same structure. For the comparison of the two, we need just the following identifications.
\begin{center}
\begin{tabular}{|c|c|}
  \hline
  PushASEP & Parallel TASEP \\
  \hline
 $\varphi(x,y)$ & $\phi^{\sharp}(x,y)$ \\
 $F_n(x,a,b)$ & $\widetilde F_n(x,t)$ \\
  \hline
\end{tabular}
\end{center}
where
\begin{equation}
\widetilde F_n(x,t)=F_n(x,t+n).
\end{equation}
From (\ref{pF3}) it then follows that $\widetilde F_n$ satisfies
\begin{equation}\label{eq3.6}
(\phi^\sharp * \widetilde F_n)(x,t)=\widetilde F_{n+1}(x,t).
\end{equation}
The identity (\ref{eq3.6}) corresponds to Lemma 4.3 of~\cite{BF07}. To obtain Proposition 3.1 of~\cite{BF07} one needs to prove Lemma 4.5 of~\cite{BF07} which needs as ingredients only Lemma 4.3 of~\cite{BF07}. These Lemmas are the only ingredients used to obtain Theorem 4.1 and then Theorem 4.2 of~\cite{BF07}. This last theorem concerns some general determinantal measures (a generalization of Lemma 3.4 in~\cite{BFPS06}), whose specialization to the PushASEP model is Proposition 3.1 of~\cite{BF07}. All these steps go through unchanged in our case provided we make the above identifications.

To obtain the precise expressions of Theorem~\ref{ThmJointCorr}, we use an integral representation for $\phi^\sharp$, namely,
\begin{equation}
\phi^{\sharp}(x,y)=\frac{1}{2\pi\I}\oint_{\Gamma_{0,-1}}\dx w \frac{1}{(w+1)^{x-y+1}}\frac{(1+pw)(1+w)}{w}.
\end{equation}
From this we get
\begin{equation}
(\phi^{\sharp})^n(x,y)=\frac{1}{2\pi\I}\oint_{\Gamma_{0,-1}}\dx w \frac{1}{(w+1)^{x-y+1}}\left(\frac{(1+pw)(1+w)}{w}\right)^n.
\end{equation}
Also, the ${\cal T}_{t_i,t_{i-1}}(x,y)$ from~\cite{BF07} now becomes
\begin{equation}
\widetilde F_0(x-y,t_i-t_{i-1})=\frac{1}{2\pi\I}\oint_{\Gamma_{0,-1}}\dx w \frac{(1+pw)^{t_i-t_{i-1}}}{(1+w)^{x-y+1}}.
\end{equation}

The only step we have still to prove is that the space where we need to do the orthogonalization is actually $V_n={\rm span}(1,x,\ldots,x^{n-1})$. This follows from Lemma~\ref{LemmaSpaceOrtho} below.
\end{proofOF}

\begin{lem}\label{LemmaSpaceOrtho}
We have the identity
\begin{equation}\label{eq3.10}
\sum_{x^k_l,1\leq l\leq k < n}
\prod_{k=1}^{n-1}\det(\phi^\sharp(x^k_i,x_j^{k+1}))_{1\leq i,j\leq k+1} = C_n \prod_{1\leq i<j\leq n} (x_j^n-x_i^n).
\end{equation}
for some constant $C_n\neq 0$.
\end{lem}
\begin{proofOF}{Lemma~\ref{LemmaSpaceOrtho}}
The proof is made by induction. Assume that (\ref{eq3.10}) is true for some $n$, which is the case for $n=2$. Then we prove that (\ref{eq3.10}) holds for $n+1$. Consider
\begin{equation}
\varphi_k(x,y)=
\begin{cases}
1-q,&x=y,\\
1, &y>x,\\
0, &y<x.
\end{cases}
\end{equation}
We will prove (\ref{eq3.10}) for $\varphi$ instead of $\phi^{\#}$; the statements are clearly equivalent by shifts $x_i^k\mapsto x_i^k+k$. Set
\begin{equation}
G^n_k(x)=\frac{1}{2\pi\I}\oint_{|z|\gg 1} \dx z \frac{(q+(1-q)z)^{n-k}}{(z-1)^{n-k+1}} z^x
\end{equation}
Then, for $x\geq 0$, $G^n_n(x)=1$ and $G^n_k(x)$ is a polynomial of degree $n-k$ in $x$ (by evaluating the residue at $z=1$), while for $x<0$, $G^n_k(x)=0$ (because the residue at $\infty$ gives zero). Therefore
\begin{equation}
\det(G^n_k(x^n_j))_{1\leq j,k\leq n} = C_n \prod_{1\leq i<j\leq n} (x_j^n-x_i^n).
\end{equation}
For the proof by induction, we apply Heine identity,
\begin{eqnarray}
& &\sum_{x_1,\ldots,x_n}\det(G^n_k(x_j))_{1\leq j,k\leq n} \det(\varphi_{n+1}(x_i,y_j))_{1\leq i,j\leq n+1} \nonumber \\ &=& n! \det((G^n_k*\varphi_{n+1})(y_j))_{1\leq j,k\leq n+1}.
\end{eqnarray}
The computation of the convolution leads to $(G^n_k*\varphi_{n+1})(y)=G^{n+1}_k(y)$.
\end{proofOF}

Theorem~\ref{ThmJointCorr} holds for general fixed initial conditions. We want to apply it to the alternating initial condition. For that we first have to do the orthogonalization with the result given in the next lemma.
\begin{lem}\label{PsiPhi}
For initial conditions $y_j=-2j$, $j=1,\ldots,n$, we have
\begin{equation}\label{Psif}
 \Psi_j^{n,t}(x)
 =
 \frac{1}{2\pi \I} \oint_{\Gamma_{0,-1}} \dx w
 \frac{w^j (1+pw)^{t-j}}{(1+w)^{x+2n-j+1}},
\end{equation}
and
\begin{equation}\label{Phif}
 \Phi_j^{n,t}(x)
 =
 \frac{1}{2\pi\I} \oint_{\Gamma_0} \dx z
 \frac{(1+2z+pz^2)(1+z)^{x+2n-j-1}}{z^{j+1}(1+pz)^{t-j+1}},
\end{equation}
where, as before, $p=1-q$. In particular, $\Phi_0^{n,t}(x)=1$.
\end{lem}

\begin{proofOF}{Lemma \ref{PsiPhi}}
The formula for $\Psi^n_j$ is just obtained by substituting the initial conditions into (\ref{eq2.4}). Now we prove that the orthonormality relation (\ref{ortho}) holds. For $k=0,\ldots,n-1$, the pole at $w=0$ in $\Psi^n_k$ is not present, and for $x<-2n+k$, $\Psi^n_k(x)=0$ because the residue at $-1$ vanishes. Thus
\begin{eqnarray}
& & \sum_{x\in\Z} \Phi_j^{n,t}(x) \Psi_k^{n,t}(x) \nonumber \\
&=& \frac{1}{(2\pi\I)^2} \oint_{\Gamma_0}\dx z \oint_{\Gamma_{-1}} \dx w
 \frac{(1+2z+pz^2)(1+z)^{2n-j-1}}{z^{j+1}(1+pz)^{t-j+1}} \frac{w^k(1+pw)^{t-k}}{(1+w)^{2n-k-1}}\nonumber \\
 & & \times \sum_{x=-2n+k}^{\infty} \left(\frac{1+z}{1+w}\right)^x
\end{eqnarray}
where we have the constraint on the integration paths $|1+z|<|1+w|$. The last term (the sum) equals
\begin{equation}
\left(\frac{1+z}{1+w}\right)^{-2n+k} \frac{1+w}{w-z}.
\end{equation}
Now the pole at $w=-1$ has disappeared and instead of it there is a simple pole at $w=z$. Thus, the integral over $w$ is just the residue at $w=z$, leading to
\begin{eqnarray}
 \sum_{x\in\Z} \Phi_j^{n,t}(x) \Psi_k^{n,t}(x) &=& \frac{1}{2\pi\I}  \oint_{\Gamma_0} \dx z \frac{1+2z + pz^2}{(1+pz)^2} \left(\frac{z(1+z)}{1+pz}\right)^{k-j-1} \nonumber \\
 &=& \frac{1}{2\pi\I}\int_{\Gamma_0} \dx u u^{k-j-1} =\delta_{j,k}
\end{eqnarray}
where we used the change of variable $u=\frac{z(1+z)}{1+pz}$.
\end{proofOF}

Lemma~\ref{PsiPhi} together with Theorem~\ref{ThmJointCorr} leads to the kernel for the alternating initial condition.
\begin{prop} \label{KNalt}
For $y_j=-2j, j=1,\ldots,n$, the kernel $K$ in Theorem~\ref{ThmJointCorr} is given by
\begin{eqnarray}
K((n_1,t_1),x_1;(n_2,t_2),x_2)& =& -\phi^{((n_1,t_1),(n_2,t_2))}(x_1,x_2) \Id_{[(n_1,t_1)\prec (n_2,t_2)]} \nonumber \\
& &+ \widetilde K((n_1,t_1),x_1;(n_2,t_2),x_2)
\end{eqnarray}
where $\phi^{((n_1,t_1),(n_2,t_2))}(x_1,x_2)$ is given by (\ref{phi}) and
\begin{align}\label{eq2.53}
\widetilde K((n_1,t_1),x_1;(n_2,t_2),x_2)
&=\frac{1}{(2\pi\I)^2} \oint_{\Gamma_{-1}} \dx w \oint_{\Gamma_0} \dx z
\frac{w^{n_1}(1+pw)^{t_1-n_1+1}}{(1+w)^{x_1+n_1+1}} \notag\\
&\quad \times
\frac{(1+z)^{x_2+n_2}(1+2z+pz^2)}{z^{n_2}(1+pz)^{t_2-n_2+2}}
\frac{1}{(w-z)(w+\frac{1+z}{1+pz})}.
\end{align}
Here $\Gamma_0$ (resp $\Gamma_{-1}$) is any simple loop, anticlockwise oriented, which includes the pole at $z=0$
(resp. $w=-1$), satisfying $\{-\frac{1+z}{1+pz},z\in\Gamma_0\} \subset \Gamma_{-1}$ and no point of $\Gamma_0$ lies inside $\Gamma_{-1}$.
\end{prop}

\begin{proofOF}{Proposition \ref{KNalt}}
We substitute (\ref{Psif}) and (\ref{Phif}) in the kernel (\ref{eqKernelFinal}). Since $\Phi^{n,t}_j(x)=0$ for $j<0$, we can extend the sum over $k$ to $\infty$. We can take the sum inside the integrals if the integration paths satisfy \mbox{$\left|\frac{1+pw}{w(1+w)}\frac{z(1+z)}{1+pz}\right|<1$}. Then we compute the geometric series and obtain
\begin{align}
\sum_{k=1}^{\infty} \Psi^{n_1,t_1}_{n_1-k}(x_1) \Phi^{n_2,t_2}_{n_2-k}(x_2)
 &= \frac{1}{(2\pi\I)^2} \oint_{\Gamma_{0,-1}} \dx w \oint_{\Gamma_0} \dx z
 \frac{w^{n_1}(1+pw)^{t_1-n_1+1}}{(1+w)^{x_1+n_1+1}} \notag\\
 &\quad \times
 \frac{(1+z)^{x_2+n_2}(1+2z+pz^2)}{z^{n_2}(1+pz)^{t_2-n_2+2}}
 \frac{1}{(w-z)(w+\frac{1+z}{1+pz})}.
\end{align}
At this point both simple poles $w=z$ and $w=-(1+z)/(1+pz)$ are inside the integration path $\Gamma_{0,-1}$, but the integrand does not have any pole anymore at $w=0$. Thus we will drop the $0$ in $\Gamma_{0,-1}$. Separating the contribution from the pole at $w=z$ we get
\begin{eqnarray}\label{eq2.55}
\sum_{k=1}^{n_2} \Psi^{n_1,t_1}_{n_1-k}(x_1) \Phi^{n_2,t_2}_{n_2-k}(x_2)& =& \widetilde K((n_1,t_1),x_1;(n_2,t_2),x_2)\\
 &+& \frac{1}{2\pi\I}\oint_{\Gamma_0} \left(\frac{1+pz}{z}\right)^{n_2-n_1}(1+z)^{n_2+x_2-n_1-x_1-1}.\nonumber
\end{eqnarray}
Moreover, we also have
\begin{eqnarray}\label{eq2.56}
\phi^{*((n_1,t_1),(n_2,t_2))}(x_1,x_2) &=& \phi^{((n_1,t_1),(n_2,t_2))}(x_1,x_2) \\
 &+& \frac{1}{2\pi\I}\oint_{\Gamma_0} \left(\frac{1+pz}{z}\right)^{n_2-n_1}(1+z)^{n_2+x_2-n_1-x_1-1}.\nonumber
\end{eqnarray}
Thus the last two terms of (\ref{eq2.55}) and (\ref{eq2.56}) cancel out, leading to (\ref{eq2.53}).
\end{proofOF}

With Proposition~\ref{KNalt} we almost obtained Theorem~\ref{ThmKalt}. What remains to do is to focus far enough into the negative axis, where the influence of the finiteness of the number of particles is not present anymore. There the kernel is equal to the kernel for the initial conditions $y_i=-2i$, $i\in\Z$.

\begin{proofOF}{Theorem \ref{ThmKalt}}
The kernel for the flat case is obtained by considering the region satisfying $x_1+n_1+1\leq 0$ where the
effect of the boundary in the TASEP is absent. Here the pole at $w=-1$ vanishes. Computing the residue at $w=-(1+z)/(1+pz)$ in Proposition \ref{KNalt} gives the kernel (\ref{Kt}) up to a factor $(-1)^{n_1-n_2}$ which we cancel by a conjugation of the kernel.
\end{proofOF}

\section{Proof of Theorem~\ref{ThmXitoAi}} \label{SectPOTXitoAi}
From Theorem~\ref{ThmKalt} we have that
$\Pb(x_n(t)\geq x)=\det(\Id-\Id_{(-\infty,x)} K \Id_{(-\infty,x)})$.
We have such a situation but with $x=y-s T^{1/3}$. With this change
of variable, we get
$\Pb(x_n(t)\geq y-s T^{1/3})=\det(\Id-\Id_{(s,\infty)}
K^{\rm resc}_T \Id_{(s,\infty)})$
where
$K^{\rm resc}_T(\xi_1,\xi_2)=T^{1/3}
K(x_1-\xi_1 T^{1/3},x_2-\xi_2 T^{1/3})$
(here we did not write explicitly the $(n,t)$ entries).
Taking into account the scaling (\ref{eqScaling1}), we thus
have to analyze the rescaled kernel
\begin{eqnarray}\label{eqRescKernel}
& &K_T^{\rm resc}(u_1,\xi_1;u_2,\xi_2)\\
&=&T^{1/3} K\big((n(u_1),t(u_1)),x(u_1)-\xi_1 T^{1/3};(n(u_2),t(u_2)),x(u_2)-\xi_2 T^{1/3}\big),\nonumber
\end{eqnarray}
with $x(u)=-2 n(u)+\mathbf{v} t(u)$, $\mathbf{v}=1-\sqrt{q}$. In particular, we have to prove that, for $u_1,u_2$ fixed, $K_T^{\rm resc}$ (or a conjugate kernel of it) converges to the kernel $\kappa_{\rm v}^{-1}K_{\Af}(\kappa_{\rm h}^{-1} u_1,\kappa_{\rm v}^{-1}\xi_1,\kappa_{\rm h}^{-1} u_2,\kappa_{\rm v}^{-1} \xi_2)$ uniformly on bounded sets and have enough control (bounds) on the decay of $K^{\rm resc}$ in the variables $\xi_1,\xi_2$ such that also the Fredholm determinant converges.

In order to have a proper limit of the kernel as $T\to\infty$, we have to consider the conjugate kernel $K_T^{\rm conj}$ given by
\begin{equation}\label{conjugation}
K_T^{\rm conj}(u_1,\xi_1;u_2,\xi_2)=K_T^{\rm resc}(u_1,\xi_1;u_2,\xi_2) \left(\frac{\sqrt{q}}{1+\sqrt{q}}\right)^{x_1-x_2} q^{n_1-n_2} q^{-(t_1-t_2)/2}.
\end{equation}
The new kernel does not change the determinantal measure, being just a
conjugation of the old one. So, in the following we will determine the limit of
$(\ref{conjugation})$ as $T\to\infty$.

\begin{prop}[Uniform convergence on compact sets]\label{PropUnifCvg}
For $u_1,u_2$ fixed, according to (\ref{eqScaling1}), set
\begin{align}
 x_i &= \left[-2n(u_i)+\mathbf{v} t(u_i)-\xi_i T^{1/3}\right], \\
 n_i &= n(u_i),\quad t_i=t(u_i).
\end{align}
Then, for any fixed $L>0$, we have
\begin{equation}
 \lim_{T\to\infty} K_T^{\rm{conj}}(n_1,x_1;n_2,x_2) T^{1/3}
 = \kappa_{\rm v}^{-1}K_{\Af}(\kappa_{\rm h}^{-1} u_1,\kappa_{\rm v}^{-1}\xi_1,\kappa_{\rm h}^{-1} u_2,\kappa_{\rm v}^{-1} \xi_2)
\end{equation}
uniformly for $(\xi_1,\xi_2)\in[-L,L]^2$, with the kernel $K_{\Af}$ given by (\ref{eqKernelExpanded}) and the constants $\kappa_{\rm v}$ and $\kappa_{\rm h}$ given by (\ref{eqkappa0}).
\end{prop}

\begin{proofOF}{Proposition \ref{PropUnifCvg}}
First we consider the first term in (\ref{K}). We thus consider (\ref{phi}) with the above replacements and conjugation. This term has to be considered only for $u_2>u_1$. The change of variable $w=-1+\sqrt{q}z$ leads then to
\begin{equation}\label{eq4.6}
\frac{T^{1/3}}{2\pi\I}\oint_{\Gamma_0} \frac{\dx z}{z} e^{T^{2/3} (g_0(z)-g_0(z_c))+T^{1/3}(g_1(z)-g_1(z_c))}
\end{equation}
with $z_c=(1+\sqrt{q})^{-1}$ and
\begin{align}
 g_0(z) &=(u_2-u_1)(\pi'(\theta)+1) \big(\ln(\sqrt{q}+(1-q)z)-(1-\sqrt{q})\ln(z)\big)\notag \\
 &+(u_2-u_1)(1-\pi'(\theta))\ln\left(\frac{\sqrt{q}+(1-q)z}{z(1-\sqrt{q}z)}\right), \notag\\
 g_1(z) &= -(u_2^2-u_1^2)\tfrac12 \pi''(\theta)\big(\sqrt{q}\ln(z)+\ln(1-\sqrt{q}z)\big)-(\xi_2-\xi_1)\ln(z).
\end{align}
The function $g_0$ has a critical point at $z=z_c$. The series expansions around $z=z_c$ are
\begin{align}\label{eq4.8}
 g_0(z) &= g_0(z_c)+(u_2-u_1) \kappa_1 (z-z_c)^2 + \Or((z-z_c)^3), \\
 g_1(z) &=g_1(z_c) -(\xi_2-\xi_1)(1+\sqrt{q})(z-z_c) + \Or((z-z_c)^2), \notag
\end{align}
where
\begin{equation}\label{eqKappa}
\kappa_1=\sqrt{q}(1+\sqrt{q})^2\left[(\pi'(\theta)+1)\frac{1-\sqrt{q}}{2}+1-\pi'(\theta)\right].
\end{equation}

To prove convergence of (\ref{eq4.6}) we have to show that the contribution coming around the critical point dominates in the $T\to\infty$ limit. We do it by finding a steep descent path\footnote{For an integral $I=\int_\gamma \dx z e^{t f(z)}$, we say that $\gamma$ is a steep descent path if (1) $\Re(f(z))$ is maximal at some $z_0\in\gamma$: $\Re(f(z))< \Re(f(z_0))$ for $z\in\gamma\setminus\{z_0\}$ and (2) $\Re(f(z))$ is monotone along $\gamma\setminus\{z_0\}$ except, if $\gamma$ is closed, at a single point where $\Re(f)$ is minimal.} for $g_0$ passing by $z=z_c$. Consider the path $\Gamma_0=\{\rho e^{i\phi},\phi\in [-\pi,\pi)\}$. Then, on $\Gamma_0$, $\frac{\dx}{\dx \phi}\Re(\ln(z))=0$,
\begin{equation}\label{eq4.50}
\frac{\dx}{\dx \phi}\Re(\ln(\sqrt{q}+(1-q)z)) = -\frac{\sqrt{q}(1-q)\rho \sin(\phi)}{|\sqrt{q}+(1-q)z|^2},
\end{equation}
and
\begin{equation}\label{eq4.51}
\frac{\dx}{\dx \phi}\Re(-\ln(1-\sqrt{q}z)) = -\frac{\sqrt{q} \rho \sin(\phi)}{|1-\sqrt{q}z|^2}.
\end{equation}
Thus $\Gamma_0$ is a steep descent path for $g_0$. Now we set $\rho=z_c$. Then, the real part of $g_0(z)$ is maximal at $z=z_c$ and strictly less then $g(z_c)$ for all other points on $\Gamma_0$. Therefore, we can restrict the integration from $\Gamma_0$ to $\Gamma_0^\delta=\{z\in\Gamma_0 | |z-z_c|\leq\delta\}$. For $\delta$ small, the error made is just of order $\Or(e^{-c T^{2/3}})$ with $c>0$ ($c\sim \delta^2$ as $\delta\ll 1$). In the integral over $\Gamma_0^\delta$ we can use (\ref{eq4.8}) to get
\begin{eqnarray}\label{eq4.10}
& &\frac{(1+\sqrt{q})T^{1/3}}{2\pi\I}\oint_{\Gamma_0^\delta} \dx z e^{T^{2/3} (u_2-u_1) \kappa_1 (z-z_c)^2
-T^{1/3}(\xi_2-\xi_1)(1+\sqrt{q})(z-z_c)}\nonumber \\
&\times & e^{\Or(T^{2/3}(z-z_c)^3,T^{1/3}(z-z_c)^2,(z-z_c))}.
\end{eqnarray}
We use $|e^x-1|\leq |x|e^{|x|}$ to control the difference between (\ref{eq4.10}) and the same expression without the error terms. By taking $\delta$ small enough and the change of variable $(z-z_c)T^{1/3}=W$, we obtain that this difference is just of order $\Or(T^{-1/3})$, uniformly for $\xi_1,\xi_2$ in a bounded set. At this point we remain with (\ref{eq4.10}) without the error terms. We extend the integration path to $z_c+\I\R$ and this, as above, gives an error of order $\Or(e^{-c T^{2/3}})$. Thus we have
\begin{eqnarray}\label{eq4.11}
(\ref{eq4.6})&=&\Or(e^{-c T^{2/3}},T^{-1/3})\\
&+&\frac{(1+\sqrt{q})T^{1/3}}{2\pi\I}\oint_{z_c+\I\R} \dx z e^{T^{2/3} (u_2-u_1) \kappa_1 (z-z_c)^2 -T^{1/3}(\xi_2-\xi_1)(1+\sqrt{q})(z-z_c)}. \nonumber
\end{eqnarray}
Therefore, uniformly for $\xi_1,\xi_2$ in bounded sets,
\begin{equation}
\lim_{T\to\infty}(\ref{eq4.6}) = \frac{1}{\sqrt{4\pi(u_2-u_1)\alpha}}\exp\left(-\frac{(\xi_2-\xi_1)^2}{4(u_2-u_1) \alpha^2}\right)
\end{equation}
with $\alpha^2=\kappa_1/(1+\sqrt{q})^2=\kappa_{\rm v}^2/\kappa_{\rm h}$.

Now we have to consider the second term in (\ref{K}). Notice that this time the restriction $u_2>u_1$ does not apply. Set $z_c=-1/(1+\sqrt{q})$. Then
\begin{equation}\label{eq4.14}
\widetilde{K}_T^{\text{conj}}(u_1,\xi_1;u_2,\xi_2)=
\frac{-T^{1/3}}{2\pi\I} \oint_{\Gamma_0} \dx z \frac{e^{T f_0(z)+ T^{2/3} f_1(z) + T^{1/3} f_2(z)+f_3(z)}}{e^{T^{2/3} f_1(z_c) + T^{1/3} f_2(z_c)}}
\end{equation}
with
\begin{align}\label{eq4.16}
 f_0(z) &= (\pi(\theta)+\theta)\Big[(1-\sqrt{q})\ln\left(\frac{1+z}{-z}\right)-(1+\sqrt{q})\ln(1+(1-q)z)+\sqrt{q}\ln(q)\Big], \notag \\
 f_1(z) &= (\pi'(\theta)+1)\Big[ u_1 ((1-\sqrt{q})\ln(-z)+\sqrt{q}\ln((1+(1-q)z)/q)) \notag\\
 &\hspace{5em} -u_2((1-\sqrt{q})\ln(1+z)-\ln(1+(1-q)z))\Big] \notag\\
 &+(1-\pi'(\theta))(u_1-u_2)\ln\left(\frac{(1+z)(-z)}{1+(1-q)z}\right),
 \end{align}
 \begin{align}\label{eq4.16b}
 f_2(z) &= \frac{\pi''(\theta)}{2}\Big[u_1^2(\ln(1+z)+\sqrt{q}\ln(-qz)-(1+\sqrt{q})\ln(1+(1-q)z))\notag \\
 &\hspace{3em}-u_2^2(\sqrt{q}\ln(1+z)+\ln(-z))\Big] \notag\\
 &+\xi_1\ln(-qz/(1+(1-q)z))-\xi_2\ln(1+z), \notag \\
 f_3(z) &= -\ln(-z(1+(1-q)z)).
\end{align}

The function $f_0$ has a double critical point at $z=z_c$ and the series expansions around $z=z_c$ of the $f_i$'s are given by
\begin{align}\label{eq4.15}
f_0(z)&=\tfrac13\kappa_2(z-z_c)^3+\Or((z-z_c)^4),\notag\\
f_1(z)&=f_1(z_c)+\frac{\kappa_1}{q}(u_2-u_1)(z-z_c)^2+\Or((z-z_c)^3),\notag \\
f_2(z)&=f_2(z_c)-(\xi_1+\xi_2)\frac{1+\sqrt{q}}{\sqrt{q}}(z-z_c)+\Or((z-z_c)^2),\notag \\
f_3(z)&=\ln((1+\sqrt{q})/\sqrt{q})+\Or((z-z_c)),
\end{align}
with $\kappa_1$ given in (\ref{eqKappa}) and
\begin{equation}\label{eqKappa2}
\kappa_2=\frac{(\pi(\theta)+\theta)(1-q)(1+\sqrt{q})^3}{q}.
\end{equation}

The leading contribution in the $T\to\infty$ limit will come from the region around the double critical point. The first step is to choose for $\gamma_0$ a steep descent path for $f_0$. First we consider $\gamma_0=\{-\rho e^{\I\phi},\phi\in[-\pi,\pi)\}$, $\rho\in (0,1/(1+\sqrt{q})]$. The only part in $\Re(f_0(z))$ which is not constant along $\gamma_0$ is the term $(\pi(\theta)+\theta)(1-\sqrt{q})$ multiplied by $A(z)=\ln|1+z|-a\ln|1+(1-q)z|$, $a=(1+\sqrt{q})/(1-\sqrt{q})$. Simple computations lead then to
\begin{equation}
\frac{\dx}{\dx \phi}A(z)= -\frac{\sin(\phi)\rho\sqrt{q}\big(2+\sqrt{q}-4\rho(1+\sqrt{q})\cos(\phi)+\rho^2(2-\sqrt{q})(1+\sqrt{q})^2\big)}{|1+z|^2 |1+(1-q)z|^2}.
\end{equation}
This expression is strictly less than zero along $\gamma_0$ except at $\phi=0,-\pi$, provided that the last term is strictly positive for $\phi\neq 0,-\pi$. This is easy to check because the last term reaches his minimum at $\cos(\phi)=-1$. Solving a second degree equations, we get that on $\rho\in(0,1/(1+\sqrt{q}))$ it is strictly positive and at $\rho=1/(1+\sqrt{q})$ is zero. Thus, the path $\gamma_0$ is steep descent for $f_0$.

But close to the critical point, the steepest descent path leaves with an angle $\pm\pi/3$. Therefore, consider for a moment $\gamma_1=\{z=z_c+e^{-\I\pi\mathrm{sgn}(x)/3}|x|,x\in \R\}$. By symmetry we can restrict the next computations to $x\geq 0$. We have to see that $B(z)=\ln|1+z|-\ln|z|-a\ln|1+(1-q)z|$ is maximum at $x=0$ and decreasing for $x>0$. We have
\begin{eqnarray}
& &\frac{\dx}{\dx x}B(z)=-\frac{x^2}{2|1+z|^2 |z|^2 |1+(1-q)z|^2}\\
&\times&\big(2q+2(1-q)\sqrt{q}x-(1-3\sqrt{q}+q)(1+\sqrt{q})^2x^2+2(1-q)(1+\sqrt{q})^3 x^3\big).\nonumber
\end{eqnarray}
The term in the second line is always positive for all $x\geq 0$. To see this, remark that it is a polynomial of third degree which goes to $\infty$ as $x\to\infty$ and at $x=0$ is already positive and has positive slope. Therefore one just computes its stationary points and, if reals, takes the right-most one. There, the term under consideration turns out to be positive, which concludes the argument. Consequently, $\gamma_1$ is also a steep descent path.

We choose a steep descent path $\Gamma_0$ as follows. We follow $\gamma_1$ starting from the critical point until we intersect it with $\gamma_0$, and then we follow $\gamma_0$. Since $\Gamma_0$ is steep descent for $f_0$, we can integrate only on $\Gamma_0^\delta=\{z\in\Gamma_0 | |z-z_c|\leq \delta\}$. The error made by this cut is just of order $\Or(e^{-c T})$ for some $c=c(\delta)>0$ (with $c\sim \delta^3$ as $\delta\to 0$). Around the critical point we use the series expansions (\ref{eq4.15}). Thus we have
\begin{eqnarray}\label{eq4.20}
& &\widetilde{K}_T^{\text{conj}}(u_1,\xi_1;u_2,\xi_2)=\Or(e^{-c T})\nonumber \\
& &+\frac{-T^{1/3}}{2\pi\I} \int_{\Gamma_0^\delta} \dx z \frac{1+\sqrt{q}}{\sqrt{q}} e^{\Or((z-z_c)^4T,(z-z_c)^3T^{2/3},(z-z_c)^2 T^{1/3},(z-z_c))}\nonumber \\
&&\times e^{\tfrac13\kappa_2(z-z_c)^3T+ \frac{\kappa_1}{q}(u_2-u_1)(z-z_c)^2T^{2/3}-(\xi_1+\xi_2)\frac{1+\sqrt{q}}{\sqrt{q}}(z-z_c)T^{1/3}}
\end{eqnarray}
We want to cancel the error terms. The difference between (\ref{eq4.20}) and the same expression without the error terms is bounded using $|e^x-1| \leq e^{|x|}|x|$, applied to $x=\Or(\cdots)$. Then, this error term becomes
\begin{eqnarray}\label{eq4.22}
& &\Big|\frac{T^{1/3}}{2\pi\I} \int_{\Gamma_0^\delta} \dx z \frac{1+\sqrt{q}}{\sqrt{q}} \Or((z-z_c)^4T,(z-z_c)^3T^{2/3},(z-z_c)^2 T^{1/3},(z-z_c))\nonumber \\
&&\times e^{c_1\tfrac13\kappa_2(z-z_c)^3T+ c_2\frac{\kappa_1}{q}(u_2-u_1)(z-z_c)^2T^{2/3}-c_3(\xi_1+\xi_2)\frac{1+\sqrt{q}}{\sqrt{q}}(z-z_c)T^{1/3}}
\Big|
\end{eqnarray}
for some $c_1,c_2,c_3$ depending on $\delta$. As $\delta\to 0$, the $c_i\to 1$. Thus, for $\delta$ small enough, we have $c_1>0$. By the change of variable $(z-z_c)T^{1/3}=W$ we obtain that (\ref{eq4.22}) is just of order $\Or(T^{-1/3})$. Thus we have
\begin{eqnarray}\label{eq4.24}
& &\widetilde{K}_T^{\text{conj}}(u_1,\xi_1;u_2,\xi_2)=\Or(e^{-c T}, T^{-1/3}) \\
& &+\frac{-T^{1/3}}{2\pi\I} \int_{\Gamma_0^\delta} \dx z \frac{1+\sqrt{q}}{\sqrt{q}} e^{\tfrac13\kappa_2(z-z_c)^3T+ \frac{\kappa_1}{q}(u_2-u_1)(z-z_c)^2T^{2/3}-(\xi_1+\xi_2)\frac{1+\sqrt{q}}{\sqrt{q}}(z-z_c)T^{1/3}}. \nonumber
\end{eqnarray}
The extension of the path $\Gamma_0^\delta$ to a path going from $e^{\I\pi/3}\infty$ to $e^{-\I\pi/3}\infty$ accounts into an error $\Or(e^{-cT})$ only. We do the change of variable $Z=\kappa_2^{1/3} T^{1/3}(z-z_c)$ and we define
\begin{equation}
\kappa_{\rm v}=\frac{\kappa_2^{1/3} \sqrt{q}}{1+\sqrt{q}},\quad \kappa_{\rm h}=\frac{\kappa_2^{2/3} q}{\kappa_1}.
\end{equation}
Then,
\begin{equation}
\lim_{T\to\infty} \widetilde{K}_T^{\text{conj}}(u_1,\xi_1;u_2,\xi_2)=\kappa_{\rm v}^{-1}\frac{-1}{2\pi\I}\int_{\gamma_\infty} \dx Z e^{\tfrac13 Z^3+(u_2-u_1)Z^2\kappa_{\rm h}^{-1}-(\xi_1+\xi_2) Z \kappa_{\rm v}^{-1}}
\end{equation}
where $\gamma_\infty$ is any path going from $e^{\I\pi/3}\infty$ to $e^{-\I\pi/3}\infty$.
The proof ends by using the Airy function representation (\ref{Airy}).
\end{proofOF}

\begin{prop}[Bound for the diffusion term of the kernel]\label{PropBoundDiffusion}
Let $n_i$, $t_i$, and $x_i$ be defined as in Proposition~\ref{PropUnifCvg}. Then, for $u_2-u_1>0$ fixed and for any $\xi_1,\xi_2\in\R$, the bound
\begin{equation}\label{eq4.30}
\Big|\phi^{((n_1,t_1),(n_2,t_2))}(x_1,x_2) T^{1/3}\left(\frac{\sqrt{q}}{1+\sqrt{q}}\right)^{x_1-x_2} q^{n_1-n_2} q^{-(t_1-t_2)/2} \Big|\leq \mathrm{const}\, e^{-|\xi_1-\xi_2|}
\end{equation}
holds for $T$ large enough and $\mathrm{const}$ independent of $T$.
\end{prop}
\begin{proofOF}{Proposition~\ref{PropBoundDiffusion}}
We start with (\ref{eq4.6}). The difference now is that the contribution coming from large $|\xi_1-\xi_2|$ can be of the same order as the one from $g_0(z)$. We consider as path $\Gamma_0=\{\rho e^{\I\phi},\phi\in[-\pi,\pi)\}$.

The difference is that now we choose $\rho$ as follows. For an $\e$ with $0<\e\ll 1$ and $z_c=1/(1+\sqrt{q})$,
\begin{equation}\label{eqRho}
\rho=
\begin{cases}
z_c+\frac{(\xi_2-\xi_1)T^{-1/3}}{(u_2-u_1)\kappa_1},& \textrm{if } |\xi_2-\xi_1|\leq \e T^{1/3},\\
z_c+\frac{\e}{(u_2-u_1)\kappa_1},& \textrm{if } \xi_2-\xi_1 \geq \e T^{1/3},\\
z_c-\frac{\e}{(u_2-u_1)\kappa_1},& \textrm{if } \xi_2-\xi_1 \leq -\e T^{1/3}.\\
\end{cases}
\end{equation}
By (\ref{eq4.50}) and (\ref{eq4.51}), $\Gamma_0$ is a steep descent path for $g_0(z)$ plus the term proportional to $\xi_1-\xi_2$ in $g_1(z)$. So, integrating on $\Gamma_0^\delta=\{z=\rho e^{\I \phi},\phi\in(-\delta,\delta)\}$ instead of $\Gamma_0$ we do only an error of order $\Or(e^{-c T^{2/3}})$ times the value at $\phi=0$, for some $c>0$. Thus
\begin{eqnarray}\label{eq4.31}
\textrm{LHS of }(\ref{eq4.30})&=& e^{T^{2/3} (g_0(\rho)-g_0(z_c))+T^{1/3}(g_1(\rho)-g_1(z_c))}\\
&\times & \Big(\Or(e^{-c T^{2/3}})+\frac{T^{1/3}}{2\pi\I}\int_{\Gamma_0^\delta}\frac{\dx z}{z} e^{T^{2/3} (g_0(z)-g_0(\rho))+T^{1/3}(g_1(z)-g_1(\rho))} \Big).\nonumber
\end{eqnarray}
On $\Gamma_0^\delta$, the $\xi_i$-dependent term in $\Re(g_1(z)-g_1(\rho))$ is equal to zero. With the same procedure as in Proposition~\ref{PropUnifCvg} one shows that the integral is bounded by a constant, uniformly in $T$.

It remains to estimate the first factor in (\ref{eq4.31}). With our choice (\ref{eqRho}), we need just series expansions of $g_0$ and $g_1$ around $\rho$. Namely, by (\ref{eq4.8})
\begin{eqnarray}\label{eq4.32}
T^{2/3} (g_0(\rho)-g_0(z_c))&=& (u_2-u_1)\kappa_1(\rho-z_c)^2T^{2/3}(1+\Or(\rho-z_c)), \nonumber \\
T^{1/3}(g_1(\rho)-g_1(z_c))&=& (\xi_1-\xi_2)(1+\sqrt{q})(\rho-z_c)T^{1/3}(1+\Or(\rho-z_c))\nonumber \\
&+&\Or((\rho-z_c)^2)T^{1/3}.
\end{eqnarray}
First consider the case $|\xi_2-\xi_1|\leq \e T^{1/3}$. We replace $\rho$ given in (\ref{eqRho}) into (\ref{eq4.32}) and get that the sum of the two contributions in (\ref{eq4.32}) writes
\begin{equation}\label{eq4.32b}
-\frac{\sqrt{q}(\xi_2-\xi_1)^2}{(u_2-u_1)\kappa_1}\big(1+\Or(\e)+\Or(T^{-1/3})\big).
\end{equation}
$\Or(\e)$ comes from $\Or(\rho-z_c)$, while the $\Or(T^{-1/3})$ from $\Or((\rho-z_c)^2)$. Then, by taking $\e$ small enough and $T$ large enough, we get
\begin{equation}
(\ref{eq4.32b}) \leq -|\xi_2-\xi_1|+\mathrm{const}.
\end{equation}
In the case, $\xi_2-\xi_1 > \e T^{1/3}$, we also replace the appropriate $\rho$ given in (\ref{eqRho}) into (\ref{eq4.32}). We explicitly use the bound $\e T^{1/3}<\xi_2-\xi_1$ to bound $\Or((\rho-z_c)^2)\leq (\xi_2-\xi_1) T^{-1/3} \e$. Then, we obtain the following bound for the sum of the two contributions in (\ref{eq4.32}),
\begin{equation}
|\xi_2-\xi_1|\e T^{1/3}\left(\Or(T^{-1/3})-\frac{\sqrt{q}}{(u_2-u_1)\kappa_1}(1+\Or(\e))\right) \leq -|\xi_2-\xi_1|
\end{equation}
by taking a fixed $\e$ small enough and then $T$ large enough. Finally, for $\xi_2-\xi_1 < \e T^{1/3}$, the same result holds in a similar way.
\end{proofOF}

\begin{prop}[Bound for the main term of the kernel]\label{PropBoundMainTerm}
Let $n_i$, $t_i$, and $x_i$ be defined as in Proposition~\ref{PropUnifCvg}. Let $L>0$ fixed. Then, for given $u_1,u_2$ and $\xi_1,\xi_2\geq -L$, the bound
\begin{equation}\label{eq4.35}
\big|\widetilde{K}_T^{\text{conj}}(u_1,\xi_1;u_2,\xi_2)\big| \leq \mathrm{const}\, e^{-(\xi_1+\xi_2)}
\end{equation}
holds for $T$ large enough and $\mathrm{const}$ independent of $T$.
\end{prop}
\begin{proofOF}{Proposition~\ref{PropBoundMainTerm}}
For $\xi_1,\xi_2\in [-L,L]$ it is the content of Proposition~\ref{PropUnifCvg}. Thus we consider $\xi_1,\xi_2\in[-L,\infty)^2\setminus[-L,L]^2$. Define $\tilde \xi_i=(\xi_i+2L) T^{-2/3}>0$. Then we consider a slight modification of (\ref{eq4.14}), namely
\begin{equation}\label{eq4.39}
\widetilde{K}_T^{\text{conj}}(u_1,\xi_1;u_2,\xi_2)=
\frac{-T^{1/3}}{2\pi\I} \oint_{\Gamma_0} \dx z \frac{e^{T \tilde f_0(z)+ T^{2/3} f_1(z) + T^{1/3} \tilde f_2(z)+f_3(z)}}{e^{T \tilde f_0(z_c)+T^{2/3} f_1(z_c) + T^{1/3} \tilde f_2(z_c)}}
\end{equation}
with $f_1(z)$ and $f_3(z)$ as in (\ref{eq4.16})-(\ref{eq4.16b}), $\tilde f_2(z)$ as $f_2(z)$ in (\ref{eq4.16b}) but with $\xi_1$ and $\xi_2$ replaced by $-2L$, and finally $\tilde f_0(z)$ is set to be equal to $f_0(z)$ in (\ref{eq4.16}) plus the term
\begin{equation}\label{eq4.40}
-\tilde \xi_1\ln((1+(1-q)z)/(-qz))-\tilde \xi_2\ln(1+z).
\end{equation}
We also chose $\Gamma_0=\{-\rho e^{\I\phi},\phi\in[-\pi,\pi)\}$. In the proof of Proposition~\ref{PropUnifCvg} we already proved that $\Gamma_0$ is a steep descent path for $f_0$ for the values $\rho\in (0,z_c]$. Also, since $\tilde \xi_i>0$, $\Re((\ref{eq4.40}))$ is also decreasing while $|\phi|$ is increasing. The precise choice of $\rho$ is
\begin{equation}\label{eqRho2}
\rho=
\begin{cases}
-z_c-((\tilde \xi_1+\tilde \xi_2)/\kappa_2)^{1/2}, &\textrm{if } |\tilde \xi_1+\tilde \xi_2|\leq \e,\\
-z_c-(\e/\kappa_2)^{1/2},&\textrm{if } |\tilde \xi_1+\tilde \xi_2|\geq \e,
\end{cases}
\end{equation}
for some small $\e>0$ which can be chosen later. Let us define
\begin{equation}
Q(\rho)=e^{\Re\big(T (\tilde f_0(-\rho)-\tilde f_0(z_c))+T^{2/3}(f_1(-\rho)-f_1(z_c))+T^{1/3}(\tilde f_2(-\rho)-\tilde f_2(z_c))\big)}.
\end{equation}
Then, since $\Gamma_0$ is a steep descent path for $\tilde f_0$,
\begin{equation}\label{eq4.42}
(\ref{eq4.39}) = Q(\rho)\Or(e^{-c T})+Q(\rho)\frac{-T^{1/3}}{2\pi\I} \int_{\Gamma_0^\delta} \dx z \frac{e^{T \tilde f_0(z)+ T^{2/3} f_1(z) + T^{1/3} \tilde f_2(z)+f_3(z)}}{e^{T \tilde f_0(\rho)+T^{2/3} f_1(\rho) + T^{1/3} \tilde f_2(\rho)}},
\end{equation}
where $\Gamma_0^\delta=\{-\rho e^{\I\phi},\phi\in (-\delta,\delta)\}$, for a small $\delta>0$. The expansion around $\phi=0$ leads to
\begin{equation}
\Re(\tilde f_0(-\rho e^{\I\phi})-\tilde f_0(-\rho))=-\gamma_1 \phi^2(1+\Or(\phi))
\end{equation}
with
\begin{eqnarray}
\gamma_1&=&\frac{(1-\rho(1+\sqrt{q}))\rho\sqrt{q}(1-\sqrt{q})(\rho q+(1-\rho)(2+\sqrt{q}))}{2(1-\rho)^2(1-\rho(1-q))^2} \nonumber \\
&+&\frac{\rho}{2}\bigg(\frac{\tilde \xi_2}{(1-\rho)^2}+\frac{\tilde \xi_1 (1-q)}{(1-\rho(1-\rho))^2}\bigg)
\end{eqnarray}
which is strictly positive for $\rho$ chosen as in (\ref{eqRho2}). Also, $\Re(f_1(-\rho e^{\I\phi})-f_1(-\rho))=\gamma_2 \phi^2(1+\Or(\phi))$ for some bounded $\gamma_2$ (we do not write it down explicitly since the precise formula is not relevant). Therefore, the last term in (\ref{eq4.42}) is bounded by
\begin{equation}\label{eq4.43}
\mathrm{const}\, Q(\rho)T^{1/3}\int_{-\delta}^\delta \dx \phi e^{-\gamma \phi^2 T(1+\Or(\phi))(1+\Or(T^{-1/3}))}
\end{equation}
with $\gamma=\gamma_1+\gamma_2 T^{-1/3}$. By choosing $\delta$ small enough and independent of $T$, and then $T$ large enough, the error terms can be replaced by $1/2$, and the integral is then bounded by the one on $\R$. Thus
\begin{equation}
(\ref{eq4.43})\leq \mathrm{const}\, Q(\rho)\frac{1}{\sqrt{\gamma T^{1/3}}}.
\end{equation}
In the worse case, when $\gamma\to 0$, which happens when $\rho\to z_c$, we have $\gamma_1 T^{1/3}\simeq(\xi_1+\xi_2+4L)^{1/2} \geq (2L)^{1/2}$, which dominates $\gamma_2$ for $L$ large enough.

Therefore we have shown that
\begin{equation}
\big|\widetilde{K}_T^{\text{conj}}(u_1,\xi_1;u_2,\xi_2)\big|\leq Q(\rho) \Or(1).
\end{equation}
It thus remains to find an bound on $Q(\rho)$. We have, by (\ref{eq4.15}),
\begin{equation}
Q(\rho)=e^{\left[\tfrac13\kappa_2(-\rho-z_c)^3-(\tilde \xi_1+\tilde \xi_2) \frac{1+\sqrt{q}}{\sqrt{q}}(-\rho-z_c)T+\frac{\kappa_1}{q}(u_2-u_1)(-\rho-z_c)^2 T^{2/3}\right](1+\Or(-\rho-z_c))}.
\end{equation}
In the case $|\tilde \xi_1+\tilde \xi_2|\leq \e$, we then obtain
\begin{equation}
Q(\rho)\leq e^{(\xi_1+\xi_2+4L)^{3/2}\big(\tfrac13-\tfrac{1+\sqrt{q}}{\sqrt{q}}\big)\kappa_2^{-1/2}(1+\Or(\e))+ (\xi_1+\xi_2+4L)\Or(1)}\leq \mathrm{const} e^{-(\xi_1+\xi_2)}
\end{equation}
for $L\gg 1$, $\e\ll 1$. Finally, when $|\tilde \xi_1+\tilde \xi_2|\geq \e$, we have
\begin{equation}
Q(\rho)\leq e^{-(\xi_1+\xi_2+4L)\big(\big(\tfrac13-\tfrac{1+\sqrt{q}}{\sqrt{q}}\big)\e^{1/2} T^{1/3}\kappa_2^{-1/2}+\Or(1)\big)}\leq e^{-(\xi_1+\xi_2)}
\end{equation}
by first choosing $\e>0$ small and then $T$ large enough.
\end{proofOF}

\begin{proofOF}{Theorem~\ref{ThmXitoAi}}
The proof of Theorem~\ref{ThmXitoAi} is the complete analogue of Theorem 2.5 in~\cite{BFP06}. The results in Propositions 5.1,5.3,5.4, and 5.5 in~\cite{BFP06} are replaced by the ones in Proposition~\ref{PropUnifCvg},~\ref{PropBoundDiffusion}, and~\ref{PropBoundMainTerm}. The strategy is to write the Fredholm series of the expression for finite $T$ and, by using the bounds in Propositions~\ref{PropBoundDiffusion} and~\ref{PropBoundMainTerm}, see that it is bounded by a $T$-independent and integrable function. Once this is proven, one can exchange the sums/integrals and the $T\to\infty$ limit by the theorem of dominated convergence. For details, see Theorem 2.5 in~\cite{BFP06}.
\end{proofOF}

\section{Proof of Theorem~\ref{ThmcPNG}} \label{SectPOTcPNG}
In this section we prove Theorem~\ref{ThmcPNG}.
By Theorem~\ref{ThmKalt}, the right hand side of (\ref{hxn})
with $n_i=\lfloor(t_i-H_i-x_i)/2\rfloor$ can be written as Fredholm
determinant of the kernel
\begin{equation}
\Id(X_i<x_i) K((n_i,t_i),X_i;(n_j,t_j),X_j) \Id(X_j<x_j)
\end{equation}
with $K$ given in (\ref{K}). By the change of variable
$X_i=-h_i+H_i+x_i$, one obtains the Fredholm determinant of the kernel
\begin{equation}
\Id(h_i>H_i) K((n_i,t_i),H_i+x_i-h_i;(n_j,t_j),H_j+x_j-h_j) \Id(h_j>H_j).
\end{equation}
With this preparation, we now go to the proof of Theorem~\ref{ThmcPNG}.

\begin{proofOF}{Theorem~\ref{ThmcPNG}}
We have to analyze the kernel (\ref{K}) with entries
\begin{equation}\label{eq5.10}
n_i=\frac{\tpng_i+\xpng_i}{\sqrt{q}}-\frac{H_i}{2},\quad t_i=\frac{2\tpng_i}{\sqrt{q}},\quad x_i=-\frac{2\xpng_i}{\sqrt{q}}+H_i-h_i
\end{equation}
and take the limit $q\to 0$ with $h_i,H_i$ fixed.
The scaling of $x_i$ might look different from the one in
(\ref{eqHeightRescaling}) but, as we can see below,
(\ref{eq5.10}) with the last one replaced by
$x_i = -\frac{2\xpng_i}{\sqrt{q}}$ gives the same limiting kernel.
As $q\to 0$, the kernel does not have a well defined limit and, as usual, we first have to consider a conjugate kernel. More precisely, we define
\begin{equation}
K_q((\xpng_1,\tpng_1),h_1;(\xpng_2,\tpng_2),h_2)=
K((n_1,t_1),x_1;(n_2,t_2),x_2)q^{(x_1-x_2)/2}\frac{q^{n_1-n_2}}{q^{(t_1-t_2)/2}}.
\end{equation}
What we have to prove is
\begin{equation}\label{eq5.5}
\lim_{q\to 0}\det(\Id-\chi_H K_q \chi_H)=\det(\Id-\chi_H K^{\rm PNG} \chi_H).
\end{equation}
First we prove the pointwise convergence and then we obtain bounds allowing us to take the limit inside the Fredholm determinant.

Consider the term coming from (\ref{phi}). By the change of variable $w=-1+\sqrt{q}z$, we get
\begin{equation}
\frac{1}{2\pi\I}\oint_{\Gamma_0}\frac{\dx z}{z} \frac{(\sqrt{q}+(1-q)z)^{t_1-t_2}}{z^{x_1-x_2}} \left(\frac{1-\sqrt{q}z}{z(\sqrt{q}+(1-q)z)}\right)^{n_1-n_2}
\end{equation}
and, by inserting (\ref{eq5.10}), one obtains
\begin{eqnarray}\label{eq5.11}
& &\frac{1}{2\pi\I}\oint_{\Gamma_0}\frac{\dx z}{z} \left(\frac{(\sqrt{q}+(1-q)z)(1-\sqrt{q}z)}{z}\right)^{(\tpng_1-\tpng_2)/\sqrt{q}}  \\
& &\left(\frac{\sqrt{q}+(1-q)z}{(1-\sqrt{q}z)z}\right)^{(\xpng_1-\xpng_2)/\sqrt{q}}
\left(\frac{\sqrt{q}+(1-q)z}{(1-\sqrt{q}z)z}\right)^{(H_1-H_2)/2}
\frac{1}{z^{h_2-h_1}}. \nonumber
\end{eqnarray}
Consider $q\leq q_0$ for some $q_0<1$ fixed. Then, we can fix the path $\Gamma_0$ independent of $q$, and the $q\to 0$ limit is easily obtained. It results in
\begin{eqnarray}
\label{PNGdiff}
& &\lim_{q\to 0}\,(\ref{eq5.11})=\frac{1}{2\pi\I}\oint_{\Gamma_0}\frac{\dx z}{z} \frac{1}{z^{h_2-h_1}} e^{-(\tpng_1-\tpng_2)(z-z^{-1})}e^{(\xpng_2-\xpng_1)(z+z^{-1})} \\
&=&\left(\frac{\xpng_2-\xpng_1+\tpng_1-\tpng_2}{\xpng_2-\xpng_1-\tpng_1+\tpng_2}\right)^{(h_1-h_2)/2} I_{h_1-h_2}\left(2\sqrt{(\xpng_2-\xpng_1)^2-(\tpng_1-\tpng_2)^2}\right), \nonumber
\end{eqnarray}
where we applied (\ref{Bes4}).

It is the turn of the term coming from (\ref{Kt}). We do the change of variable $z=-w/(w+\sqrt{q})$ and then we insert (\ref{eq5.10}). The result is
\begin{eqnarray}\label{eq5.12}
& &\frac{1}{2\pi\I}\oint_{\Gamma_0}\frac{\dx w}{w(1+\sqrt{q}w)}\left(\frac{\sqrt{q}+w}{w(1+\sqrt{q}w)}\right)^{(\tpng_1+\tpng_2)/\sqrt{q}}
\left(\frac{w}{1+\sqrt{q}w}\right)^{h_1}(w+\sqrt{q})^{h_2}\nonumber \\
& & \left(\frac{w}{(w+\sqrt{q})(1+\sqrt{q}w)}\right)^{(\xpng_1-\xpng_2)/\sqrt{q}+(H_2-H_1)/2}.
\end{eqnarray}
If $\xpng_2-\xpng_1>\tpng_1+\tpng_2$, then for $q$ small enough, the result is identically equal to zero, because the pole at $w=0$. If $\xpng_1-\xpng_2>\tpng_1+\tpng_2$, then the result is also zero, because the residues at all other poles, $\sqrt{q}$, $1/\sqrt{q}$, and $\infty$ vanishes. In the other case, when $|\xpng_2-\xpng_1|<\tpng_1+\tpng_2$, the apparent pole at $w=-\sqrt{q}$ is actually not there. So, we can choose a $\Gamma_0$ independent of $q\leq q_0$ for some $q_0<1$. Then, we can simply take the limit $q\to 0$ of the integrand, which leads to
\begin{eqnarray}
\label{PNGmain}
& &\lim_{q\to 0}\, (\ref{eq5.12}) = \frac{1}{2\pi\I}\oint_{\Gamma_0}\frac{\dx w}{w} w^{h_1+h_2} e^{(\tpng_1+\tpng_2)(w^{-1}-w)}e^{(\xpng_2-\xpng_1)(w+w^{-1})}\\
&=&\left(\frac{\tpng_1+\tpng_2+\xpng_2-\xpng_1}{\tpng_1+\tpng_2-\xpng_2+\xpng_1}\right)^{(h_1+h_2)/2} J_{h_1+h_2}\left(2\sqrt{(\tpng_1+\tpng_2)^2-(\xpng_2-\xpng_1)^2}\right),\nonumber
\end{eqnarray}
where in the last step we made the change of variable $w\to 1/w$ and applied (\ref{Bes3}).

To have convergence of the Fredholm determinants we still need some bounds for large values of $h_1,h_2$. For $q$ small enough, say $q\in [0,q_0]$ for some $q_0<1$, we can set in (\ref{eq5.11}) $\Gamma_0=\{z, |z|=\mathrm{e}\}$ in the case $h_2\geq h_1$, and $\Gamma_0=\{z, |z|=\mathrm{e}^{-1}\}$ in the case $h_2<h_1$. Then, we get the bound
\begin{equation}
|(\ref{eq5.11})|\leq C_1 e^{-|h_2-h_1|}
\end{equation}
for some finite constant $C_1$ independent of $q$. Moreover, in (\ref{eq5.12}) we can choose $\Gamma_0=\{z, |z|=\mathrm{e}^{-2}\}$, which leads to the bound
\begin{equation}
|(\ref{eq5.12})|\leq C_2 e^{-(h_2+h_1)}
\end{equation}
with $C_2<\infty$ independent of $q\leq q_0$. These two bounds are enough to have convergence of the Fredholm determinants. The strategy is exactly the same as in the proof of Theorem~\ref{ThmXitoAi}.
\end{proofOF}

\section{Proof of Theorem~\ref{ThmXiTAi}} \label{SectPOTXiTAi}

We analyze the kernel (\ref{PNGkernel}) with the scalings
\begin{align}
 \xpng_i &= u_i T^{2/3}, \\
 \tpng_i &= \gamma(0)T + \gamma'(0)u_i T^{2/3}
           +\frac{\gamma''(0)}{2}u_i^2 T^{1/3}, \\
 h_i     &= 2\tpng_i + \xi_i T^{1/3}
\end{align}
(See (\ref{eq1.24}) and (\ref{XPt})).
The strategy of he proof is the same as that for Theorem
\ref{ThmXitoAi} and hence we only give the main differences.

First we consider the first term in (\ref{PNGkernel}). From
(\ref{PNGdiff}) it is rewritten in the from (\ref{eq4.6}) with
$g_0(z),g_1(z)$ replaced by
\begin{align}
 g_0(z) &= (u_2-u_1) \left(\gamma'(0)(z-1/z-2\ln z)+(z+1/z)\right), \\
 g_1(z) &= \frac{\gamma''(0)}{2}(u_2^2-u_1^2)(z-1/z-2\ln z)
          -(\xi_2-\xi_1)\ln z.
\end{align}
The critical point of $g_0(z)$ is now $z_c=1$.
The series expansions around $z=z_c$ are
\begin{align}
 g_0(z) &= g_0(z_c) + (u_2-u_1)(z-z_c)^2+ O((z-z_c)^3), \\
 g_1(z) &= g_1(z_c) -(\xi_2-\xi_1)(z-z_c) + O((z-z_c)^2).
\end{align}
The steep descent path can be taken to be
$\Gamma_0 = \{ e^{i\phi}, \phi\in[-\pi,\pi)\}$.
Then the same arguments as in the proof of Theorem \ref{ThmXitoAi}
give the first term in (\ref{eqKernelExpanded}).

Next we consider the second term in (\ref{PNGkernel}).
From (\ref{PNGmain}) it is rewritten in the form (\ref{eq4.14})
with $f_0(z),f_1(z),f_2(z)$ replaced by
\begin{align}
 f_0(w) &=2\gamma(0) (1/w-w+2\ln w), \\
 f_1(w) &= \gamma'(0)(u_1+u_2) (1/w-w+2\ln w) + (u_2-u_1)(w+1/w), \\
 f_2(w) &= \frac{\gamma''(0)}{2}(u_1^2+u_2^2)(1/w-w+2\ln w)
           +(\xi_1+\xi_2)\ln w.
\end{align}
Their series expansions around $z_c$ are
\begin{align}
 f_0(w) &= -\frac{2\gamma(0)}{3}(w-1)^3+ O((w-1)^4), \\
 f_1(w) &= f_1(z_c) + (u_2-u_1)(w-1)^2+ O((w-1)^3), \\
 f_2(w) &= f_2(z_c) + (\xi_1+\xi_2)(w-1) + O((w-1)^2).
\end{align}
The steepest descent path is taken to be
$\Gamma_0 = \{ \rho e^{i\phi}, \phi\in[-\pi,\pi)\}$ with $0<\rho<1$.
Using these one obtains the second term in (\ref{eqKernelExpanded}).

The bounds for the diffusion terms and the main term of the kernel
are also proved in the same way as those of Propositions
\ref{PropBoundDiffusion} and \ref{PropBoundMainTerm}.

\appendix
\section{Some integral representations}\label{AppBessel}
In this appendix we list some integral representations of the Bessel functions and the modified Bessel functions (we use the conventions of~\cite{AS84}).
\begin{equation}\label{Bes1}
J_n(2t)=\frac{1}{2\pi\I}\oint_{\Gamma_0}\frac{\dx z}{z}
        \frac{e^{t(z-z^{-1})}}{z^n},
\end{equation}

\begin{equation}\label{Bes2}
I_n(2t)=\frac{1}{2\pi\I}\oint_{\Gamma_0}\frac{\dx z}{z}
        \frac{e^{t(z+z^{-1})}}{z^n},
\end{equation}

\begin{equation}\label{Bes3}
\frac{1}{2\pi\I}\oint_{\Gamma_0}\frac{\dx z}{z}\frac{e^{b(z-z^{-1})}e^{a(z+z^{-1})}}{z^n} =\left(\frac{b+a}{b-a}\right)^{n/2} J_n\left(2\sqrt{b^2-a^2}\right),
\end{equation}

\begin{equation}\label{Bes4}
\frac{1}{2\pi\I}\oint_{\Gamma_0}\frac{\dx z}{z}\frac{e^{b(z-z^{-1})}e^{a(z+z^{-1})}}{z^n} =\left(\frac{a+b}{a-b}\right)^{n/2} I_n\left(2\sqrt{a^2-b^2}\right),
\end{equation}

\begin{equation}\label{Airy}
\frac{-1}{2\pi \I}\int_{\gamma_\infty}\dx v e^{v^3/3+av^2+bv}=\Ai(a^2-b)\exp(2a^3/3-ab),
\end{equation}
where $\gamma_\infty$ is any path from $e^{\I\pi/3}\infty$ to $e^{-\I\pi/3}\infty$.

\end{document}